# On the design of fixed-gain tracking filters by pole placement:

# Or an introduction to applied signals-and-systems theory for engineers

Hugh L. Kennedy
ISS Division, DST Group, Hugh.Kennedy@dst.defence.gov.au
STEM, University of South Australia, Hugh.Kennedy@unisa.edu.au
Technical Knockout Systems, Hugh.TKS@gmail.com
Adelaide, Australia

## Abstract

The Kalman filter computes the optimal variable-gain using prior knowledge of the initial state and random (process and measurement) noise distributions, which are assumed to be Gaussian with known variance. However, when these distributions are unknown, the Kalman filter is not necessarily optimal and other simpler state-estimators, such as fixed-gain ($\alpha$, $\alpha - \beta$ or $\alpha - \beta - \gamma$ etc.) filters may be sufficient. When such filters are used as low-complexity state-estimators in embedded tracking systems, the fixed gain parameters are usually set equal to the steady-state gains of the corresponding Kalman filter. An alternative procedure, that does not rely prior distributions, based on Luenberger observers, is presented here. It is suggested that the arbitrary placement of closed-loop state-observer poles is a simple and intuitive way of tuning the transient and steady-state response of a fixed-gain tracking filter when prior distributions are unknown. All poles are placed inside the unit circle on the positive real axis of the complex $z$-plane at $p$ for a well damped response and a configurable bandwidth. Transient bias errors, e.g. due to target manoeuvres or process modelling errors, decrease as $p = 0$ is approached for a wider bandwidth. Steady-state random errors, e.g. due to sensor noise, decrease as $p = 1$ is approached for a narrower bandwidth. Thus the $p$ parameter (with $0 < p < 1$) may be interpreted as a dimensionless smoothing factor. This tutorial-style report examines state-observer design by pole placement, which is a standard procedure for feedback controls but unusual for tracking filters, due to the success and popularity of the Kalman filter. As Bayesian trackers are designed via statistical modelling, not by pole-zero placement in the complex plane, the underlying principles of linear time-invariant signals and systems are also reviewed.

# Contents

# List of abbreviations and acronyms

| | |
|---|---|
| 1-D | one dimensional |
| 2-D | two dimensional |
| CCF or ccf | controllable canonical form |
| dc | direct current (i.e. 0 Hz) |
| DFT | discrete Fourier transform |
| DSP | digital signal processing |
| FIR | finite-impulse-response |
| IIR | infinite-impulse-response |
| KIN or kin | kinematic |
| LDE | linear-difference equation |
| LSS | linear state-space |
| LTI | linear time-invariant |
| MIMO | multiple-input/multiple-output |
| obs | observer |
| OCF or ocf | observable canonical form |
| PCF or pcf | process canonical form |
| prc | process |
| prd | predictor/prediction |
| SIMO | single-input/multiple-output |
| SISO | single-input/single-output |
| WNG | white-noise gain |
| w.r.t | with respect to |

# Basic mathematical definitions and symbols

$i = \sqrt{-1}$: Complex unit.
$s = \sigma + \Omega i$: Complex $s$-plane coordinate, reached via the Laplace transform.
$\sigma$: Real part of $s$ (reciprocal seconds).
$\Omega$: Imaginary part of $s$, angular frequency (radians per second).
$\tau = 1/\sigma$: Coherence duration (seconds).
$\lambda = 2\pi/\Omega$: Wave period (seconds).
$z$: Complex $z$-plane coordinate, reached via the $\mathcal{Z}$ transform.
$\omega = \Omega/F_s$: Normalized angular frequency (radians per sample).
$f = \omega/2\pi$: Normalized frequency (cycles per sample).
$F$: Frequency (cycles per second or Hz).
$F_s$: Sampling frequency i.e. sampling rate (cycles per second or Hz).
$T_s = 1/F_s$: Sampling period (seconds).
$t$: Time (seconds).
$n$: Time index, into a sampled sequence (samples, $0 \leq n < N$, $t = nT_s$).
$m$: Delay index, into a sample history (samples, $0 \leq m < M$, $t = nT_s - mT_s$).
$q$: Group delay parameter (samples, $-\infty < q < +\infty$, $t = nT_s - qT_s$).
$p$: Pole position in the complex $z$-plane.
$k$: Basis function, state vector, or operator, index ($0 \leq k < K$).
$\mathbf{A}$, $\mathbf{B}$ & $\mathbf{C}$: Continuous-time LSS matrices of an LTI system.
$\mathcal{H}(s)$: Continuous-time transfer-function of an LTI system.
$H(\Omega)$ or $H(F)$: Continuous-time frequency-response of an LTI system.
$h(t)$: Continuous-time impulse-response of an LTI system.
$\mathbf{G}$, $\mathbf{H}$ & $\mathbf{C}$: Discrete-time LSS matrices of an LTI system.
$\mathcal{H}(z)$: Discrete-time transfer-function of an LTI system.
$H(\omega)$ or $H(f)$: Discrete-time frequency-response of an LTI system.
$h[m]$: Discrete-time impulse-response of an LTI system.
$x[n]$: System input.
$y[n]$: System output.
$\mathbf{w}[n]$: System internal state vector.
$\mathcal{K}$: Observer gain vector.
$\mathbf{b}$ & $\mathbf{a}$: Coefficients of numerator and denominator polynomials in $\mathcal{H}(z) = \mathcal{B}(z)/\mathcal{A}(z)$.
$(\blacksquare)$: Denotes a function of continuous argument.
$[\blacksquare]$: Denotes a sampled function of integer argument.
$|\blacksquare|$: Magnitude of a complex variable or the determinant of a matrix.
$\angle\blacksquare$: Angle of a complex variable, e.g. $\omega = \angle z$.
$\text{Re}(\blacksquare)$: Real part of a complex variable.
$\text{Im}(\blacksquare)$: Imaginary part of a complex variable, e.g. $\Omega = \text{Im}(s)$.
$\lfloor\blacksquare\rfloor$: Rounds down to the nearest integer.
$\blacksquare^T$: Transpose of a real matrix or vector.
$\blacksquare!$: Factorial operator.

# 1. Introduction

The first-order ($\alpha$-) filter, with a single pole on the positive real axis of the complex $z$-plane, is the simplest of all infinite impulse-response (IIR) filters and it is widely used to recursively compute moving (exponentially-weighted) averages of uniformly sampled time series. However, a second-order ($\alpha - \beta$) filter should be considered when a wider bandwidth is required for smoothing and differentiating signals produced by non-stationary processes that have poles near the origin of the complex $s$-plane; or a third-order ($\alpha - \beta - \gamma$) filter if second derivatives (w.r.t. time) are approximately constant over many samples.

Higher order tracking filters highlight the need for a principled approach to the determination of gain coefficients (i.e. $\alpha$, $\beta$, $\gamma$ etc.) for a satisfactory transient and steady-state response. The steady-state gains of the corresponding Kalman filter are usually used for this purpose and it has been shown that, for second- and third-order integrating systems, the gain coefficients are a function of the tracking index $\lambda_{\text{trk}} = T_s^2 \, \sigma_Q / \sigma_R$ where $T_s$ is the sampling period and $\sigma_Q^2$ & $\sigma_R^2$ are the variance of the process noise and measurement noise, which are assumed to be Gaussian [1],[2],[3],[4],[5]. More general procedures are also described in [1] & [6]. When the noise is white, but non-Gaussian, and the variance is unknown, such procedures are convenient, however the computed gains are no longer optimal and the response of the filter is not necessarily better in a given application and environment than those found via other methods, e.g. heuristically, empirically, or manually.

The fact that the gain vector may also be derived by 'pole placement' appears to have been overlooked in the target-tracking literature. This standard procedure is usually used to design Luenberger observers for state-space feedback controls [7], where robust performance in the presence of model uncertainty is essential. The purpose of this tutorial is to introduce this old technique to a new audience.

An observer incorporates a discrete-time model of a process, with unknown internal states to be estimated. The observer and the process are both linear state-space (LSS) systems, that incorporate feedback, thus their behaviour (at the discrete sample times, $t = nT_s$, for $n = 0 \dots \infty$) is determined by the locations of their system poles (in the complex $z$-plane). The poles of the process are governed by the laws of nature, e.g. Newton's laws of motion; however, the poles of the observer are defined by the intent of the engineer and they may be placed arbitrarily, by setting the coefficients of the gain vector appropriately, to accommodate both the dynamics of the process and the requirements of the observer. The proposed tracker design procedure assumes that the engineer has a basic understanding of linear time-invariant (LTI) system modelling and an appreciation of how pole locations (in the $s$- and $z$-domains) determine the response of a system (in the $t$- and $n$-domains) [8],[9]; thus, this background material is summarized in an appendix (see Section 12).

For context, this tutorial begins with a discussion of feedback systems and the rationale behind the use Luenberger observers instead of Kalman filters in Section 2. Linear state-space models of natural processes in continuous time are then introduced in Section 3; followed by linear state-space representations of state observers in discrete time in Section

4. Observer design by pole placement is then presented in Sections 5 & 6 via a sequence of coordinate transforms; a worked example is provided in an appendix (see Section 11). Further coordinate transformations required for minimum complexity realizations, gain vector construction (i.e. $\alpha$, $\beta$ & $\gamma$), filter coefficient extraction (i.e. **a** & **b**), and response analysis, are then described in Section 7. This is followed by a summary of some standard techniques for filter response analysis and an illustration of the way in which $p$ parameter affects the (transient and steady-state) response in Section 8.

## 2.  Recursive state estimation as digital feedback design

The sequence of detection, tracking and control is an information-processing design-pattern at the core of tactical and strategic defence-systems alike. In this tutorial, an attempt is made to unify and simplify these components by focusing on the theory and practice of linear LTI systems, with an emphasis on feedback. Feedback is a powerful but unpredictable force that defies intuition. It is both a workhorse to be harnessed and a beast to be avoided. It bestows machines with a deftness that rivals the skill of a human; for instance, the ailerons of a jet/missile as it approaches a runway/target. But it also causes public-announcement systems to deafen crowds at concerts and brings cars off dirt roads as panicking drivers oversteer then understeer with increasing fervour. Naïvely feeding the output of a stable (natural or synthetic) system back into its input rarely results in a favourable outcome. However, simply shifting the phase and the scaling the gain at appropriate frequencies at a single point somewhere inside the loop is sufficient, to render stability in an amplifier (as shown by Bode) and impart apparent intelligence to a machine (as shown by Weiner), so the system output quickly converges on the desired endpoint, with minimal overshoot, rapid settling, and partial noise immunity. The early work of the feedback pioneers was all done with analogue electronics (during world-war II) and an understanding of feedback was the only way to automate machinery (e.g. fire-control systems). In modern times, with digital computers responsible almost exclusively for automation, the laissez faire approach to feedback in analogue electronics has been replaced by a more interventionary approach in digital electronics, whereby alternative action policies or processing algorithms are adopted for different system states and operating regimes. Such systems are difficult to test in isolation, and federations of such systems are difficult to manage. Designing complex systems-of-systems around the praxis of linear time-invariance (LTI) was unavoidable in older analogue systems, i.e. where feedback does the work, but it is optional in newer digital systems. The behaviour of an LTI system is predictable, deterministic, and invariant – all inputs are treated equally, which greatly reduces test coverage and integration effort. It could be that its omission from more recent large-system designs is a contributing factor in the delayed delivery and performance short-falls that are now all-too common in the development and procurement of new (civilian and defence) capabilities involving a high degree of automation. Feedback is a powerful technique for architecting, analysing, designing, and implementing, complex systems requiring machine intelligence, e.g. the detection, tracking, and control, pipeline behind an imaging sensor. Alternative batch-based LTI solutions are feasible (e.g. using the fast Fourier transform) for spatial processing at the front end of the processing pipeline; however, they become less feasible for temporal processing where low latencies are a priority. Thus, the emphasis here is on feedback systems. Extra effort is required to understand the theory that accounts of the behaviour of such systems; however, the potential rewards are great.

In this preliminary section the overlap between tracking, control, and DSP, is explored. The state-estimation theory in this tutorial is applicable to all three areas of digital system design, as it may be used to make state estimators for target trackers, state observers for feedback controls, or define the internal states of IIR digital filters for detecting, smoothing, interpolating, extrapolating, or differentiating the time varying intensity of a pixel in an image.

For the state-estimation problem, the Luenberger observer is a standard 'control' solution [10],[11]; whereas the Kalman filter is a standard 'tracking' solution [12]. It is therefore reasonable to ask: "What then is the difference between control and tracking?" In both cases the aim is to *infer* the unknown state of an entity (referred to as the 'plant' and the 'target' in control and tracking, respectively) from a sequence of sensor samples in the presence random measurement errors and model uncertainty. But in the former case the loop is closed, and the aim is to also *affect* and *change* the state of that, usually non-compliant and un-cooperative, entity. Thus, the performance of the state estimator in a controller is judged on the response of the closed-loop system as whole, incorporating both the inner observer loop and the outer actuator loop (see Figure 1). The primary input to this closed-loop system is the *desired* state of the plant and the output is the *actual* state of the plant. Whereas, the performance of a tracker is simply based on its open-loop state-estimation error.

This difference between controls and trackers gives rise to different design priorities and techniques. In control, guaranteed stability, and reasonable performance, over *all* operating conditions (i.e. robustness), is given priority over precision and accuracy in ideal operating conditions [13],[14]. In tracking, precision and accuracy are favoured, and it is achieved by leveraging prior assumptions regarding *expected* operating conditions, e.g. using a Kalman filter [12]. However, the distinction between tracker and observer is subtle and not always clear, particularly in modern defence systems, where the degree of connectivity and automation is high for very fast observation/actuation cycles. For instance: Is a tracker or an observer used as the state estimator in the seeker of a (active or passive) homing missile? And is a tracker or an observer used as the state estimator in a surveillance radar used by an operator with a communication link to a stealthy interceptor aircraft?

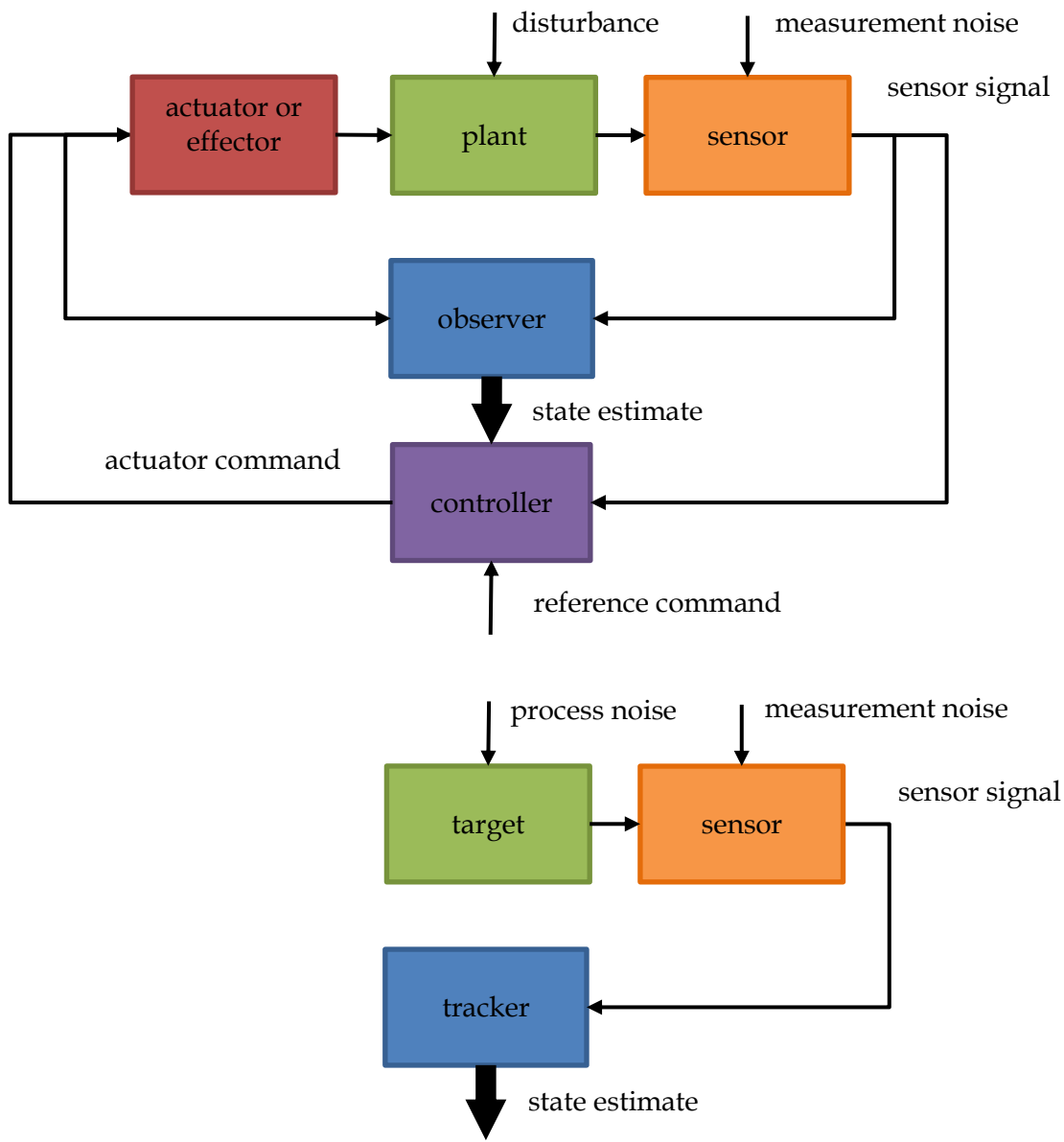

*Figure 1 – State estimation: in (closed-loop) control, e.g. using a Luenberger observer, with deterministic and random inputs (top); and in (open-loop) tracking, e.g. using a Kalman filter, with random inputs (bottom). In the control case, the transfer function linking any of the inputs to an output at any internal point of the loop, e.g. an element of the state-estimate vector, the actuator command, or the sensor signal, may be derived. Thus, the response to an arbitrary input (stochastic or deterministic) may be computed. However, there are usually too few degrees of freedom (in the observer and controller) and too many process model approximations (in the actuator, plant, and sensor) to meet all requirements simultaneously with confidence. A stable response is essential, i.e. one that produces a bounded output for a bounded input, but a zero tracking-error is preferable, and a well-damped transient-response is desirable.*

It is suggested here that all state estimators in modern (electronic) defence systems are operating in a closed-loop context even when there is a supervisory operator involved. And as such, robust behaviour (i.e. stability and performance) is always a priority. Particularly when they are expected to operate effectively and reliably in harsh, dangerous, and uncertain environments. For these reasons, the linear Luenberger observer [10],[11], which is usually used to design simple feedback controls, via pole placement, is considered here for online tracking problems that would usually be solved using the linear Kalman filter [12].

A Bayesian recursion yields the Kalman filter when Gaussian priors are assumed, and particularly simple forms are reached for linear process models and constant sampling rates. They have a variable gain that is computed by propagating the first and second moments (i.e. mean and variance) of the posterior distributions. The gain coefficients are data independent, therefore they may be pre-computed offline, and they quickly converge on limiting values in high-bandwidth systems. In this steady-state regime, the Kalman filter has the same form as a Luenberger observer. Therefore, the coefficients in the gain vector of an observer with a Luenberger structure, which does not (usually) consider second moments, may be determined from a steady-state Kalman filter when the parameters of prior distributions are known. When these parameters are unknown, the coefficients of reasonable estimators are readily obtained from the simple application of transform theory (see Section 12), to the design of digital feedback loops, using the methods discussed below.

In a Luenberger observer, the poles of the loop are placed to impart stability and the desired bandwidth for a satisfactory balance between the transient response and the steady-state response (see Figure 2). The response may be tuned to be optimal for a given set of Gaussian noise parameters, using the steady-state gain of a Kalman filter, or it may be tuned to give a reasonable response for an ensemble of representative inputs of stochastic and/or deterministic nature. The procedure is described in the sections that follow, starting with the process model, then the way in which this model is incorporated into the observer structure, with the desired response. This is followed by realization and parameterization considerations.

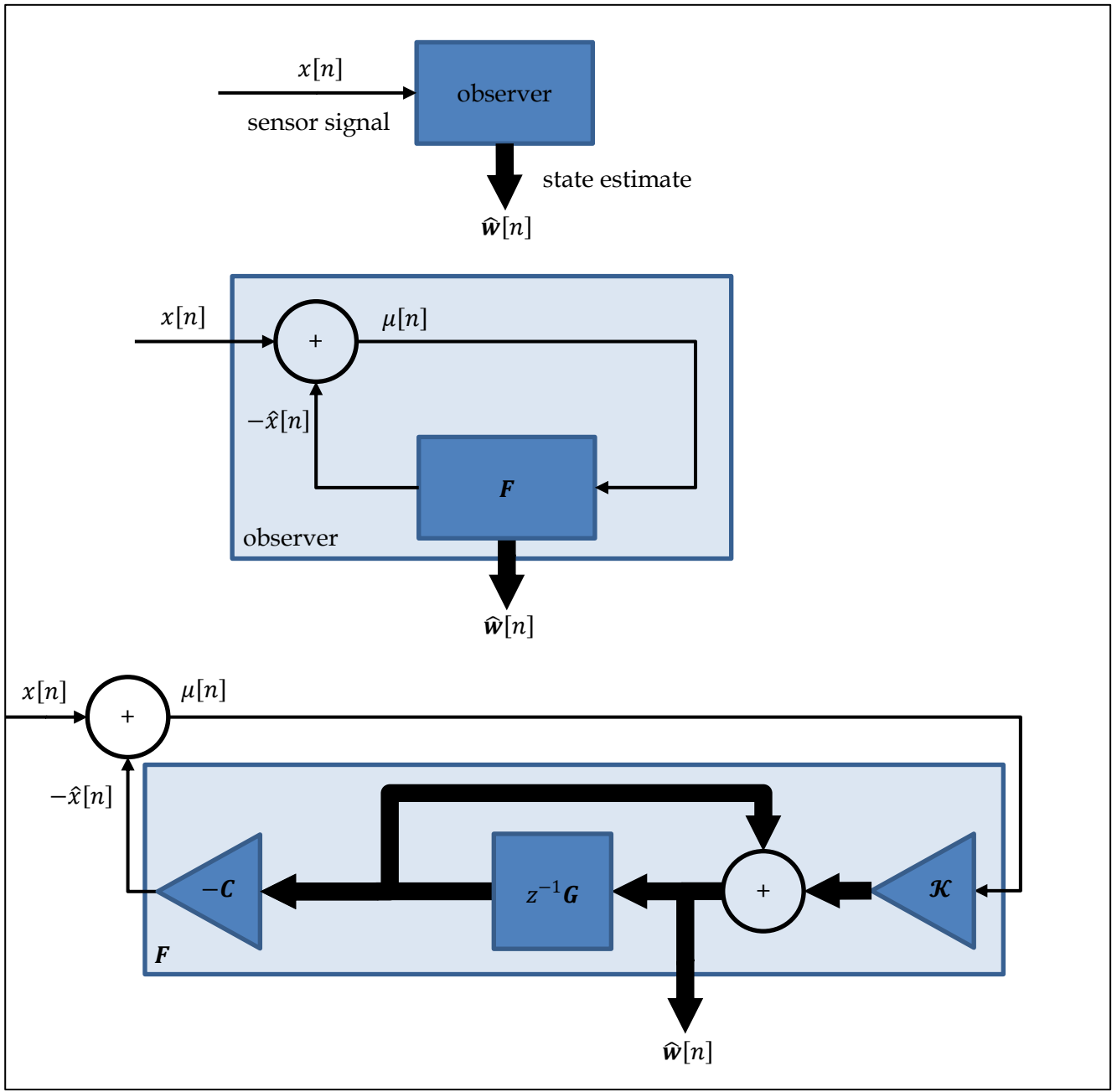

*Figure 2 – 'Unboxing' the state estimator: An observer as a tracker, i.e. with only one input (top). The observer is driven by an error signal, which is the actual measurement minus the predicted measurement, i.e. the so-called 'innovation' (middle). The observer uses a model of the process, as defined by state-transition matrix **G**, and the measurement row-vector **C**, to compute the predicted measurement (bottom). The gain column-vector **𝒦**, is used to stabilize the loop and to impart the desired transient and steady state properties of the estimator. Its coefficients are determined by a pole placement procedure.*

# 3. Process state-equations

For a tracking filter that estimates temporal derivatives (e.g. an $\alpha - \beta$ or $\alpha - \beta - \gamma$ filter) the signal is assumed to be generated by a deterministic $K$th-order integrating process, i.e. with $K$ repeated poles at $s = 0$ in the complex $s$-plane. The continuous-time LSS representation of this process, in observable canonical form (OCF) [11], is as follows:

$$\dot{\boldsymbol{w}}_{\text{prc}}(t) = \boldsymbol{A}_{\text{prc}}\boldsymbol{w}_{\text{prc}}(t) \text{ and} \tag{3.1a}$$

$$y_{\text{prc}}(t) = \boldsymbol{C}_{\text{prc}}\boldsymbol{w}_{\text{prc}}(t) \text{ with} \tag{3.1b}$$

$$\boldsymbol{A}_{\text{prc}} = \begin{bmatrix} \boldsymbol{0}_{(K-1)\times 1} & \boldsymbol{I}_{(K-1)\times(K-1)} \\ 0 & \boldsymbol{0}_{1\times(K-1)} \end{bmatrix}_{K\times K} \text{ and} \tag{3.1c}$$

$$\boldsymbol{C}_{\text{prc}} = [1 \quad \boldsymbol{0}_{1\times(K-1)}]_{1\times K} \tag{3.1d}$$

where $\boldsymbol{I}_{M\times N}$ is an $M \times N$ identity matrix and $\boldsymbol{0}_{M\times N}$ is an $M \times N$ matrix of zeros. Thus $\boldsymbol{A}_{\text{prc}}$ is a zero matrix with ones along the 1st upper diagonal, i.e.

$$\boldsymbol{A}_{\text{prc}} = \begin{bmatrix} 0 & 1 & 0 & \cdots & 0 & 0 & 0 \\ 0 & 0 & 1 & \cdots & 0 & 0 & 0 \\ 0 & 0 & 0 & \cdots & 0 & 0 & 0 \\ \vdots & \vdots & \vdots & \ddots & \vdots & \vdots & \vdots \\ 0 & 0 & 0 & \cdots & 0 & 1 & 0 \\ 0 & 0 & 0 & \cdots & 0 & 0 & 1 \\ 0 & 0 & 0 & \cdots & 0 & 0 & 0 \end{bmatrix}_{K\times K}. \tag{3.1e}$$

The output of this process is equal to the 0th element of the $K \times 1$ state vector $\boldsymbol{w}_{\text{prc}}(t)$, thus the $k$th element (for $k = 0 \ldots K - 1$) is the $k$th temporal derivative of the output at time $t$. The system is assumed to be a deterministic endogenous process (i.e. with no inputs) so that all future outputs are determined by the current internal state.

The signal model is discretized by taking the Laplace transform ($t \to s$) of (3.1a), using an initial state of $\boldsymbol{w}_{\text{prc}}(0)$

$$s\boldsymbol{W}_{\text{prc}}(s) - \boldsymbol{w}_{\text{prc}}(0) = \boldsymbol{A}_{\text{prc}}\boldsymbol{W}_{\text{prc}}(s) \tag{3.2a}$$

$$\{s\boldsymbol{I} - \boldsymbol{A}_{\text{prc}}\}\boldsymbol{W}_{\text{prc}}(s) = \boldsymbol{w}_{\text{prc}}(0) \tag{3.2b}$$

$$\boldsymbol{W}_{\text{prc}}(s) = \{s\boldsymbol{I}_{K_{\text{prc}}} - \boldsymbol{A}_{\text{prc}}\}^{-1}\boldsymbol{w}_{\text{prc}}(0). \text{ Let} \tag{3.2c}$$

$$\boldsymbol{\Phi}_{\text{prc}}(s) = \{s\boldsymbol{I}_{K_{\text{prc}}} - \boldsymbol{A}_{\text{prc}}\}^{-1}. \tag{3.2d}$$

Thus, the so-called 'fundamental matrix' $\boldsymbol{\Phi}_{\text{prc}}(t)$,
is found by taking the inverse Laplace transform ($t \leftarrow s$) of $\boldsymbol{\Phi}_{\text{prc}}(s)$

$$\boldsymbol{\Phi}_{\text{prc}}(t) = \mathcal{L}^{-1}\{\boldsymbol{\Phi}_{\text{prc}}(s)\}. \tag{3.2e}$$

The state-transition matrix $\boldsymbol{G}_{\text{prc}}^{\text{kin}}$ in 'kinematic' coordinates,
for a constant sampling period of $T_s$, is

$$\boldsymbol{G}_{\text{prc}}^{\text{kin}} = \boldsymbol{\Phi}_{\text{prc}}(t)\big|_{t=T_s} \tag{3.2f}$$

where $\boldsymbol{G}_{\text{prc}}^{\text{kin}}$ is an upper triangular Toeplitz matrix with the elements along the $k$th off-diagonal (for $k = 0 \ldots K - 1$) equal to

$$\mathcal{G}(k; T_s) = \frac{1}{(K-1)!} \frac{d^l}{dt^l} t^{K-1} \bigg|_{t=T_s} = \frac{T_s^k}{k!} \tag{3.2g}$$

where $l = K - k - 1$, i.e.

$$\boldsymbol{G}_{\text{prc}}^{\text{kin}} = \begin{bmatrix} 1 & T_s & \cdots & \frac{T_s^k}{k!} & \cdots & \frac{T_s^{(K-2)}}{(K-2)!} & \frac{T_s^{(K-1)}}{(K-1)!} \\ 0 & 1 & \cdots & \frac{T_s^{(k-1)}}{(k-1)!} & \cdots & \frac{T_s^{(K-3)}}{(K-3)!} & \frac{T_s^{(K-2)}}{(K-2)!} \\ \vdots & \vdots & \ddots & \vdots & \ddots & \vdots & \vdots \\ 0 & 0 & \cdots & 1 & \cdots & \frac{T_s^{(k-1)}}{(k-1)!} & \frac{T_s^k}{k!} \\ \vdots & \vdots & \ddots & \vdots & \ddots & \vdots & \vdots \\ 0 & 0 & \cdots & 0 & \cdots & 1 & T_s \\ 0 & 0 & \cdots & 0 & \cdots & 0 & 1 \end{bmatrix}_{K \times K}. \tag{3.2h}$$

As the diagonal elements of this triangular matrix are equal to unity, the discrete signal model has $K$ repeated poles at $z = 1$, for a singularity at dc, i.e. at $\omega = 0$, where $\omega$ is the angular frequency (radians per sample).

The discrete-time model of the process that generates the sampled signal (i.e. 'the process') may now be defined as

$$\boldsymbol{w}_{\text{prc}}^{\text{kin}}[n] = \boldsymbol{G}_{\text{prc}}^{\text{kin}} \boldsymbol{w}_{\text{prc}}^{\text{kin}}[n-1] \text{ and} \tag{3.3a}$$
$$y_{\text{prc}}[n] = \boldsymbol{C}_{\text{prc}}^{\text{kin}} \boldsymbol{w}_{\text{prc}}^{\text{kin}}[n] \text{ where} \tag{3.3b}$$
$$\boldsymbol{C}_{\text{prc}}^{\text{kin}} = \boldsymbol{C}_{\text{prc}}. \tag{3.3c}$$

The superscript is used to indicate that this formulation is for a state vector in kinematic coordinates.

Integrating models with derivative state are often used to approximate unknown processes over short time scales, in the same way that Taylor series expansions are used to approximate high-order phenomena. Alternative signal and noise models are considered elsewhere [17],[18].

# 4. Observer state-equations

As the process has $K$ poles on the unit circle, it is marginally stable. We now seek a stable observer (i.e. all poles inside the unit circle) that is cascaded with the process with the following discrete-time LSS representation (see Figure 3 and Figure 4):

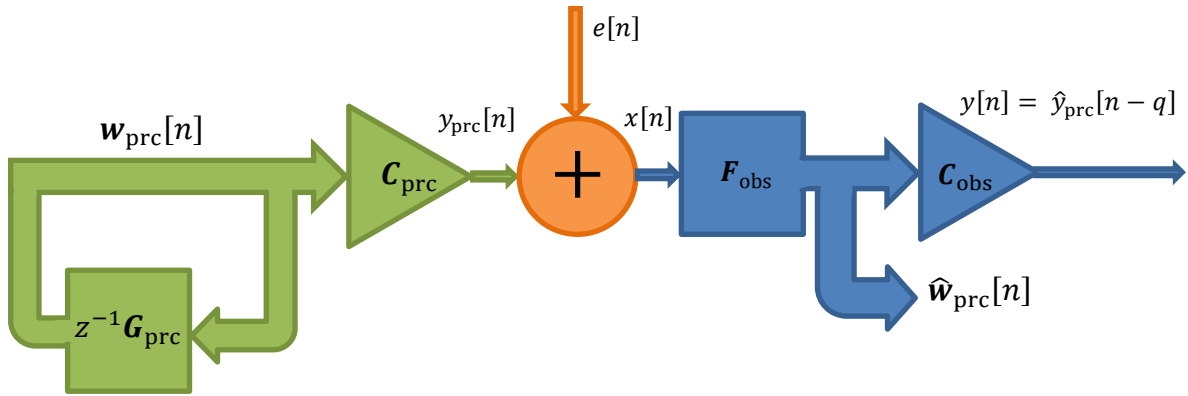

*Figure 3 -  The deterministic process system (green), as represented in (3.3), connected to the digital observer system (blue). The 'process' is a (sampled) natural phenomenon occurring in the environment; the 'observer' is a synthetic (i.e. man-made) machine, operating in a computer. Measurement error $e[n]$ with zero mean and variance of $\sigma_R^2$ that is not necessarily Gaussian, is added between the process and the observer. Thick and thin arrows are for vector- and scalar-valued connections, respectively. Square and triangular blocks represent matrix and vector operations, respectively. See Figure 4 for an expanded view of the observer.*

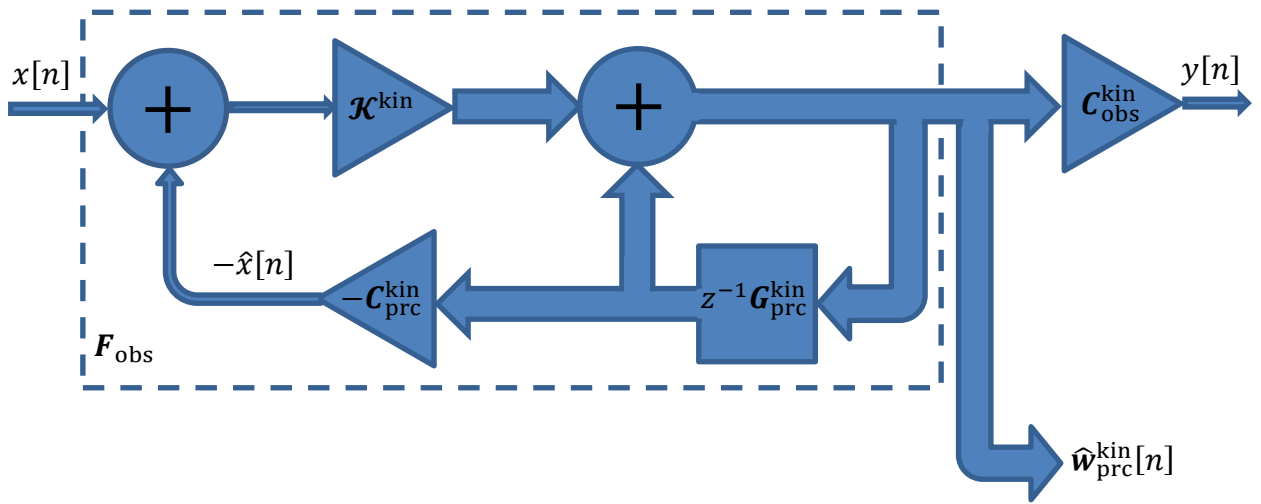

*Figure 4 -  Block diagram for the ($\mathcal{Z}$-transformed) observer system, as represented in (4.1). This is an expanded view of the observer in Figure 3 and a rearrangement/extension of Figure 2.*

In the above figures (Figure 3 and Figure 4) the following relationships and definitions are used:

$\widehat{\boldsymbol{w}}_{\text{prc}}^{\text{kin}}[n] = \boldsymbol{G}_{\text{prc}}^{\text{kin}}\widehat{\boldsymbol{w}}_{\text{prc}}^{\text{kin}}[n-1] + \boldsymbol{\mathcal{K}}^{\text{kin}}\{x[n] - \hat{x}[n]\}$ and
$y[n] = \boldsymbol{C}_{\text{obs}}^{\text{kin}}\widehat{\boldsymbol{w}}_{\text{prc}}^{\text{kin}}(n)$ where
$y[n]$ is the output, which is an estimate of $y_{\text{prc}}[n-q]$
$\boldsymbol{C}_{\text{obs}}^{\text{kin}} = \boldsymbol{C}_{\text{prc}}\boldsymbol{G}_{\text{prc}}^{\text{kin}}\{q\}$
$\widehat{\boldsymbol{w}}_{\text{prc}}^{\text{kin}}[n]$ is the estimate of $\boldsymbol{w}_{\text{prc}}^{\text{kin}}[n]$ at the time of the $n$th sample, i.e. $t = nT_s$ (seconds)
$\boldsymbol{\mathcal{K}}^{\text{kin}}$ is the $K \times 1$ observer gain vector
$\hat{x}[n]$ is the predicted input, which is an estimate of $y_{\text{prc}}[n]$
$x[n]$ is the observer input, with $x[n] = y_{\text{prc}}[n]$
$z^{-1}$ is the unit delay of one sample period ($T_s$, seconds). (4.1)

In the above definitions $q$ is the delay parameter (in samples) and $\boldsymbol{G}_{\text{prc}}^{\text{kin}}\{q\}$ is the state transition matrix for a time-step of $-qT_s$ seconds, i.e.

$$\boldsymbol{G}_{\text{prc}}^{\text{kin}}\{q\} = \boldsymbol{\Phi}_{\text{prc}}(t)\big|_{t=-qT_s}. \tag{4.2}$$

When $q$ is an integer, $\boldsymbol{G}\{q\} = \boldsymbol{G}^{-q}$ and $\boldsymbol{G}\{-q\} = \boldsymbol{G}^q$, where $\boldsymbol{G}^q$ is $q$ consecutive multiplications of $\boldsymbol{I}$ by $\boldsymbol{G}$, i.e.

$$\boldsymbol{G}^q = \underbrace{\boldsymbol{G}\ldots\boldsymbol{G}}_{q} \text{ and } \boldsymbol{G}^{-q} = \underbrace{\boldsymbol{G}^{-1}\ldots\boldsymbol{G}^{-1}}_{q}. \tag{4.3}$$

Furthermore, the structure of $\boldsymbol{G}_{\text{prc}}^{\text{kin}}$ (or $\boldsymbol{G}_{\text{prc}}$, with the superscript dropped) means that $\boldsymbol{G}_{\text{prc}}^{-1}$ has the same Toeplitz structure as $\boldsymbol{G}_{\text{prc}}$; thus, $\boldsymbol{G}_{\text{prc}}^{-1}$ is simply populated using $\mathcal{G}(k; -T_s)$, for $k = 0\ldots K-1$. Thus $\boldsymbol{G}_{\text{prc}}^{\text{kin}}\{q\}$ is a 'prediction' operation for $q < 0$ and a 'retrodiction' operation for $q > 0$.

The observer above is a single-input/multiple-output (SIMO) system, as it produces an estimate of the full kinematic state-vector with $K$ state derivatives. It also outputs an estimate of the position (i.e. the $k = 0$ element of the state vector) at a time that is $q$ samples in the past ($q > 0$), present ($q = 0$), or future ($q < 0$). When the full state vector is ignored and only used internally to produce the position, the *observer* reduces to a single-input/single-output (SISO) system, or a *smoother*, with impulse response $h[n]$, transfer function $\mathcal{H}(z)$, and frequency response $H(\omega)$ – all in discrete time. In principle, any state element, i.e. the $k$the derivative of position w.r.t. time (for $0 \leq k < K$) may be selected. For instance, a $k$th-order *differentiator* is obtained using $\boldsymbol{C}_{\text{obs}}^{\text{kin}} = \boldsymbol{C}_{\text{drv}}\boldsymbol{G}_{\text{prc}}^{\text{kin}}\{q\}$ where $\boldsymbol{C}_{\text{drv}}$ is a $1 \times K$ vector of zeros, with the $k$th element equal to unity.

The time delay (or advance) applied to the output is a design parameter; however, an advance of one sample is required to generate the error signal that 'drives' the observer. The predicted input, derived using all preceding inputs, is evaluated using

$$\hat{x}[n] = \boldsymbol{C}_{\text{prd}}^{\text{kin}}\widehat{\boldsymbol{w}}_{\text{prc}}^{\text{kin}}[n-1], \text{ where} \tag{4.4a}$$

$$\boldsymbol{C}_{\text{prd}}^{\text{kin}} = \boldsymbol{C}_{\text{prc}}^{\text{kin}} \boldsymbol{G}_{\text{prc}}^{\text{kin}}. \tag{4.4b}$$

The elements of the gain vector $\boldsymbol{\mathcal{K}}^{\text{kin}}$ are arbitrarily chosen to place the poles of the observer in the complex z-plane for the desired convergence characteristics. This gain vector may be interpreted as the extent to which the error signal, or so-called 'innovation', i.e. $\mu[n] = x[n] - \hat{x}[n]$, shifts the various elements of the state on each update. Let the $k$th (complex) pole of the observer be $\lambda_k$ (for $0 \leq k < K$). All poles must be placed in a way that results in a stable feedback observer (i.e. $|\lambda_k| < 1$). It is also desirable to have a system that is not too oscillatory (i.e. $\angle \lambda_k \approx 0$), with a rate of decay that is slow enough to attenuate white noise (i.e. $|\lambda_k| \to 1$), yet fast enough to quickly remove bias arising from discontinuities in the signal state and from system modelling errors (i.e. $|\lambda_k| \to 0$). For a complex argument (arg), |arg| and ∠arg are the magnitude and angle operators, respectively. Using repeated real poles, i.e. $\lambda_k = p$, for all $k$, with $0 \leq p < 1$, should be sufficient in most cases. The $p$ parameter may be interpreted as a smoothing factor and set using $p = e^{-1/l}$ where l is the observer 'memory' in samples ($l \geq 0$). Use $p \to 1$ (long memory) for a narrow bandwidth, high bias and low variance. Use $p \to 0$ (short memory) for a wide bandwidth, low bias and high variance. Responses for various combinations of $p$ & $q$ are presented in Section 8.

With the desired position of the observer poles in mind, some algebraic manipulations are required to determine $\boldsymbol{\mathcal{K}}^{\text{kin}}$. This begins with the substitution of (4.4) into (4.1) then rearranging, for

$$\widehat{\boldsymbol{w}}_{\text{prc}}^{\text{kin}}[n] = \{\boldsymbol{G}_{\text{prc}}^{\text{kin}} - \boldsymbol{\mathcal{K}}^{\text{kin}} \boldsymbol{C}_{\text{prd}}^{\text{kin}}\} \widehat{\boldsymbol{w}}_{\text{prc}}^{\text{kin}}[n-1] + \boldsymbol{\mathcal{K}}^{\text{kin}} x[n]. \tag{4.5}$$

The LSS representation of the observer in kinematic coordinates is now

$$\boldsymbol{w}_{\text{obs}}^{\text{kin}}[n] = \boldsymbol{G}_{\text{obs}}^{\text{kin}} \boldsymbol{w}_{\text{obs}}^{\text{kin}}[n-1] + \boldsymbol{H}_{\text{obs}}^{\text{kin}} x[n] \text{ and} \tag{4.6a}$$
$$y[n] = \boldsymbol{C}_{\text{obs}}^{\text{kin}} \boldsymbol{w}_{\text{obs}}^{\text{kin}}[n] \text{ where} \tag{4.6b}$$
$$\boldsymbol{G}_{\text{obs}}^{\text{kin}} = \boldsymbol{G}_{\text{prc}}^{\text{kin}} - \boldsymbol{\mathcal{K}}^{\text{kin}} \boldsymbol{C}_{\text{prd}}^{\text{kin}}. \tag{4.6c}$$
$$\boldsymbol{H}_{\text{obs}}^{\text{kin}} = \boldsymbol{\mathcal{K}}^{\text{kin}} \text{ and} \tag{4.6d}$$
$$\boldsymbol{w}_{\text{obs}}^{\text{kin}} = \widehat{\boldsymbol{w}}_{\text{prc}}^{\text{kin}} \tag{4.6e}$$
$$\boldsymbol{w}_{\text{obs}}^{\text{kin}}[0] = \{\boldsymbol{C}_{\text{prc}}^{\text{kin}}\}^{\text{T}} x[0] \text{ for initialization,} \tag{4.6f}$$
where the T superscript is the transpose operator.

Expressing a linear time-invariant (LTI) system, i.e. a natural sampled process or a synthetic digital observer, using LSS equations of the form

$$\boldsymbol{w}[n] = \boldsymbol{G}\boldsymbol{w}[n-1] + \boldsymbol{H}x[n] \tag{4.7a}$$
$$y[n] = \boldsymbol{C}\boldsymbol{w}[n] \tag{4.7b}$$

is somewhat unconventional. The following form is more commonly used when describing discrete-time LTI systems, in control problems [10]:

$$\boldsymbol{w}[n+1] = \boldsymbol{G}\boldsymbol{w}[n] + \boldsymbol{H}x[n] \tag{4.8a}$$
$$y[n] = \boldsymbol{C}\boldsymbol{w}[n] + \boldsymbol{D}x[n] \tag{4.8b}$$

possibly because it is analogous to the standard form for continuous-time LTI systems [11]:

$$\dot{\boldsymbol{w}}(t) = \boldsymbol{A}\boldsymbol{w}(t) + \boldsymbol{B}x(t) \quad (4.9\text{a})$$
$$y(t) = \boldsymbol{C}\boldsymbol{w}(t) + \boldsymbol{D}x(t). \quad (4.9\text{b})$$

Using (4.7) means that an input may pass immediately through the system, without delay; whereas a non-zero $\boldsymbol{D}$ matrix, is required when (4.8) is used. The inclusion of the $\boldsymbol{D}$ operator complicates analysis and design somewhat. However, it does allow an extra degree of freedom to be modelled, via an additional system zero. Although, this is not required for the systems considered here. Furthermore, designing observers via (4.8) leads to the so-called 'prediction' observer [10], where the state at the $n$th sample does not use the measurement at that time. Starting with an observer system in the form of (4.7) is well suited to the problem of state estimation because it means that additional working is not required to reach the so-called 'current' observer [10], which is the form used in other state estimators, such as the Kalman filter and the $\alpha - \beta(-\gamma)$ filter. Note that the use of $x$, $\boldsymbol{w}$ and $y$ is also uncommon in this context; it is used here to be consistent with DSP terminology [15].

# 5. Determination of system poles

Before attempting to determine $\mathcal{K}$, it is necessary to understand how the poles of the system in (4.7) are computed [10]. We begin by incrementing all indices by a forward shift of one sample so that we will be dealing with positive powers of $z$.

(Note that $z$ is a unit *advance* operator and $z^{-1}$ is a unit *delay* operator. For causal realizations, all powers of $z$ must be non-positive so that future inputs/outputs are not required to compute the current output. However, positive powers may be used for system design and analysis; for instance, it is easier to identify system poles and zeros for polynomials with positive powers.)

$$w[n+1] = Gw[n] + Hx[n+1] \tag{5.1a}$$
$$y[n+1] = Cw[n+1]. \tag{5.1b}$$

The $\mathcal{Z}$ transform of all terms is then applied, assuming a zero initial-state

$$zW(z) = GW(z) + zHX(z) \tag{5.2a}$$
$$zY(z) = zCW(z) \text{ or } Y(z) = CW(z). \tag{5.2b}$$

Then rearranging (5.2a)

$$zW(z) - GW(z) = \{zH\}X(z) \tag{5.3a}$$
$$\{zI_K - G\}W(z) = \{zH\}X(z) \tag{5.3b}$$
$$W(z) = \{zI_K - G\}^{-1}\{zH\}X(z). \text{ Let} \tag{5.3c}$$
$$\Phi(z) = \{zI_K - G\}^{-1} \text{ so now} \tag{5.3d}$$
$$W(z) = \Phi(z)\{zH\}X(z). \tag{5.3e}$$

After substituting (5.3e) into (5.2b)
$$Y(z) = C\Phi(z)\{zH\}X(z) \tag{5.4a}$$
then dividing both sides by $X(z)$
$$Y(z)/X(z) = C\Phi(z)\{zH\} \text{ or} \tag{5.4b}$$
$$\mathcal{H}^{y \leftarrow x}(z) = C\Phi(z)\{zH\} \text{ where} \tag{5.4c}$$
$$\mathcal{H}^{y \leftarrow x}(z) = Y(z)/X(z) \tag{5.4d}$$
is the discrete-time transfer-function (by definition) which links the $\mathcal{Z}$ transform of the output, i.e. $Y(z) = \mathcal{Z}\{y[n]\}$, to the $\mathcal{Z}$ transform of the input, i.e. $X(z) = \mathcal{Z}\{x[n]\}$.

The inverse in (5.3d) may be written in terms of the adjoint of $\{zI_K - G\}$ [10], using

$$\{zI_K - G\}^{-1} = \frac{\text{adj}(zI_K - G)}{|zI_K - G|}. \tag{5.5}$$

Computing the adjoint and determinants in the numerator and denominator, for matrices that are symbolic in $z$, are time-consuming tasks and fortunately, the inverse is not required here; however, this analysis does show that the poles of $\mathcal{H}(z)$ are equal to the roots of the denominator polynomial, i.e. the solutions of

$$|z\mathbf{I} - \mathbf{G}| = 0. \tag{5.6a}$$

This yields the so-called characteristic equation of the LSS system

$$z^K + \sum_{k=1}^{K} a[k] z^{K-k} = 0 \tag{5.6b}$$

which is solved for $\mathbf{a}$ by finding the eigenvalues of $\mathbf{G}$, where $\mathbf{a}$ is a vector of length $K+1$, indexed as $0 \leq k \leq K$, containing the coefficients $a[k]$, with $a[0] = 1$.

# 6. Placement of observer poles

Substitution of $G_{\text{obs}}^{\text{kin}}$ in (4.6c) for $G$ in (5.6a) yields a $z$ polynomial in (5.6b) with $a$ coefficients containing the scrambled elements of $\mathcal{K}^{\text{kin}}$. The indecipherable form follows from the fact that $\mathcal{K}^{\text{kin}} C_{\text{prd}}^{\text{kin}}$ in (4.6c) is (usually) a full $K \times K$ matrix. It is therefore not clear how the elements of $\mathcal{K}^{\text{kin}}$ should be chosen in this coordinate system to place the observer poles at the desired locations. To address this problem a new coordinate system is used [11]. The new *canonical* coordinates are a linear combination of the old *kinematic* coordinates, i.e.

$$w_{\text{obs}}^{\text{pcf}} = \mathbb{T}_{\text{prc}}^{\text{pcf} \leftarrow \text{kin}} w_{\text{prc}}^{\text{kin}}; \text{ thus, in (4.6) above} \tag{6.1a}$$

$$G_{\text{obs}}^{\text{kin}} = G_{\text{prc}}^{\text{kin}} - \mathcal{K}^{\text{kin}} C_{\text{prd}}^{\text{kin}} \text{ and } H_{\text{obs}}^{\text{kin}} = \mathcal{K}^{\text{kin}} \text{ becomes} \tag{6.1b}$$

$$G_{\text{obs}}^{\text{pcf}} = G_{\text{prc}}^{\text{pcf}} - \mathcal{K}^{\text{pcf}} C_{\text{prd}}^{\text{pcf}} \text{ and } H_{\text{obs}}^{\text{pcf}} = \mathcal{K}^{\text{pcf}} \text{ where} \tag{6.1c}$$

$\mathcal{K}^{\text{kin}}$ is the gain vector in the (old) kinematic coordinate system and
$\mathcal{K}^{\text{pcf}}$ is the gain vector in the (new) canonical coordinate system.
(Superscripts are used to identify the coordinate system.)

The linear transformation $\mathbb{T}_{\text{prc}}^{\text{pcf} \leftarrow \text{kin}}$ preserves the eigenvalues (i.e. the poles) of $G_{\text{prc}}^{\text{kin}}$ and it is specifically formulated so that it reduces the $\langle C_{\text{prd}}^{\text{kin}}, G_{\text{prc}}^{\text{kin}} \rangle$ observable pair to what is referred to here as *process* canonical form (PCF) which is (by definition):

$$G_{\text{prc}}^{\text{pcf}} = \begin{bmatrix} 0 & 0 & \cdots & 0 & 0 & g_{\text{prc}}[0] \\ 1 & 0 & \cdots & 0 & 0 & g_{\text{prc}}[1] \\ 0 & 1 & \cdots & 0 & 0 & g_{\text{prc}}[2] \\ \vdots & \vdots & \ddots & \vdots & \vdots & \vdots \\ 0 & 0 & \cdots & 1 & 0 & g_{\text{prc}}[K-2] \\ 0 & 0 & \cdots & 0 & 1 & g_{\text{prc}}[K-1] \end{bmatrix}_{K \times K} = \begin{bmatrix} \mathbf{0}_{1 \times (K-1)} & g_{\text{prc}} \\ I_{(K-1)} & \end{bmatrix}_{K \times K} \tag{6.2a}$$

$$C_{\text{prd}}^{\text{pcf}} = [0 \ 0 \ 0 \ \cdots \ 0 \ 1]_{1 \times K}. \tag{6.2b}$$

This form is ideal because the eigenvalues of $G_{\text{prc}}^{\text{pcf}}$ are readily extracted from its $K$th column, as shown below in (6.6). This form is also ideal because only the $K$th element of $C_{\text{prd}}^{\text{pcf}}$ is non-zero. The latter feature of this canonical form ensures that $\mathcal{K}^{\text{pcf}}$ only appears in the $K$th column of $G_{\text{obs}}^{\text{pcf}}$. This means that $\mathcal{K}^{\text{pcf}}$ may now be used to determine the response of the observer. The transient response is important because it determines how the observer responds to abrupt changes in the signal states, e.g. brief manoeuvre (in target-tracking systems) or edges (in computer-vision systems). The steady-state (i.e. frequency) response is important because it determines how the observer responds to sustained constant-g (i.e. circular) turns and white noise at steady state [17],[18]. Using $p \rightarrow 0$ improves the transient response (i.e. decreases bias); whereas $p \rightarrow 1$ improves the steady-state response (i.e. decreases variance); for example, using $p = 0.8$ was found to be a reasonable compromise for the simulated scenario in [17], various values of the $p$ parameter are compared in the extended treatment provided in [18]. Note that the $p$ parameter used here is the same as the $\theta$ parameter used in the fading-memory filters of [25], which are derived using discounted least-squares regression with the orthogonal (discrete) Laguerre polynomials as the regressors.

In PCF, the state-transition matrix of the observer in (6.1b) becomes

$$\boldsymbol{G}_{\text{obs}}^{\text{pcf}} = \begin{bmatrix} 0 & 0 & \cdots & 0 & 0 & g_{\text{prc}}[0] - \mathcal{K}^{\text{pcf}}[0] \\ 1 & 0 & \cdots & 0 & 0 & g_{\text{prc}}[1] - \mathcal{K}^{\text{pcf}}[1] \\ 0 & 1 & \cdots & 0 & 0 & g_{\text{prc}}[2] - \mathcal{K}^{\text{pcf}}[2] \\ \vdots & \vdots & \ddots & \vdots & \vdots & \vdots \\ 0 & 0 & \cdots & 1 & 0 & g_{\text{prc}}[K-2] - \mathcal{K}^{\text{pcf}}[K-2] \\ 0 & 0 & \cdots & 0 & 1 & g_{\text{prc}}[K-1] - \mathcal{K}^{\text{pcf}}[K-1] \end{bmatrix}_{K \times K} =$$

$$\begin{bmatrix} \boldsymbol{0}_{1 \times (K-1)} & \boldsymbol{g}_{\text{prc}} - \boldsymbol{\mathcal{K}}^{\text{pcf}} \\ \boldsymbol{I}_{(K-1) \times (K-1)} & \end{bmatrix}_{K \times K} . \tag{6.3a}$$

Now let
$$g_{\text{obs}}[k] = g_{\text{prc}}[k] - \mathcal{K}^{\text{pcf}}[k] \text{ or} \tag{6.3b}$$
$$\boldsymbol{g}_{\text{obs}} = \boldsymbol{g}_{\text{prc}} - \boldsymbol{\mathcal{K}}^{\text{pcf}} \text{ so that} \tag{6.3c}$$

$$\boldsymbol{G}_{\text{obs}}^{\text{pcf}} = \begin{bmatrix} 0 & 0 & \cdots & 0 & 0 & g_{\text{obs}}[0] \\ 1 & 0 & \cdots & 0 & 0 & g_{\text{obs}}[1] \\ 0 & 1 & \cdots & 0 & 0 & g_{\text{obs}}[2] \\ \vdots & \vdots & \ddots & \vdots & \vdots & \vdots \\ 0 & 0 & \cdots & 1 & 0 & g_{\text{obs}}[K-2] \\ 0 & 0 & \cdots & 0 & 1 & g_{\text{obs}}[K-1] \end{bmatrix}_{K \times K} = \begin{bmatrix} \boldsymbol{0}_{1 \times (K-1)} & \boldsymbol{g}_{\text{obs}} \\ \boldsymbol{I}_{(K-1) \times (K-1)} & \end{bmatrix}_{K \times K} . \tag{6.4}$$

In PCF, the effect that $\boldsymbol{\mathcal{K}}^{\text{pcf}}$ has on the poles of $\boldsymbol{G}_{\text{obs}}^{\text{pcf}}$ is now apparent. For a matrix with the canonical structure of $\boldsymbol{G}_{\text{obs}}^{\text{pcf}}$ (or $\boldsymbol{G}_{\text{prc}}^{\text{pcf}}$), the eigenvalues are equal to the roots of a polynomial formed from its $K$th column as follows:

$$z^K - \sum_{k=0}^{K-1} g_{\text{obs}}(k) z^k = \prod_{k=0}^{K-1}(z - \lambda_k). \tag{6.5}$$

Let the discrete-time transfer function relating the input $x$, to the output $y$, i.e. $\mathcal{H}(z)$ be the ratio of two polynomials $\mathcal{B}_{\text{obs}}(z)$ and $\mathcal{A}_{\text{obs}}(z)$ that determine the zeros and poles of the observer, respectively, i.e. $\mathcal{H}(z) = \mathcal{B}_{\text{obs}}(z)/\mathcal{A}_{\text{obs}}(z)$. As the eigenvalues of $\boldsymbol{G}_{\text{obs}}^{\text{pcf}}$ are equal to the poles of the observer, we have

$$\mathcal{A}_{\text{obs}}(z) = z^K - \sum_{k=0}^{K-1} g_{\text{obs}}[k] z^k = z^K + \sum_{k=1}^{K} a_{\text{obs}}[k] z^{K-k} \text{ thus} \tag{6.6a}$$
$$a_{\text{obs}}[k] = -g_{\text{obs}}[K-k], \text{ for } 1 \leq k \leq K \text{ or} \tag{6.6b}$$
$$g_{\text{obs}}[k] = -a_{\text{obs}}[K-k], \text{ for } 0 \leq k < K. \tag{6.6c}$$
i.e. opposite sign with indexing reversed and offset by one place.

In PCF it is apparent that: The elements of $\boldsymbol{\mathcal{K}}^{\text{pcf}}$ are simply chosen to yield an observer polynomial $\mathcal{A}_{\text{obs}}(z)$, that has roots at the desired pole locations $\lambda_k$, (= $p$); given the process polynomial $\mathcal{A}_{\text{prc}}(z)$, with roots at the pole locations $\rho_k$, where $|\rho_k| = 1$ for $0 \leq k < K$, as specified in the process model in (3.3). Thus, the gain vector is determined using

$$\mathcal{K}^{\text{pcf}} = \boldsymbol{g}_{\text{prc}} - \boldsymbol{g}_{\text{obs}} \text{ where} \tag{6.7}$$

$\boldsymbol{g}_{\text{prc}}$ is extracted from $\boldsymbol{G}^{\text{pcf}}_{\text{prc}}$ as shown in (6.2a) and
$\boldsymbol{g}_{\text{obs}}$ is extracted from $\boldsymbol{G}^{\text{pcf}}_{\text{obs}}$ as shown in (6.4).

The poles of the observer are set for the desired tracking behaviour – for transient inputs and at steady state for sustained inputs. The elements of $\boldsymbol{g}_{\text{obs}}$ are obtained from the poles of the observer as follows:

$$\begin{aligned}
\mathcal{A}_{\text{obs}}(z) &= \prod_{k=0}^{K-1}(z - \lambda_k) \\
&= z^K + a_{\text{obs}}[1]z^{K-1} + a_{\text{obs}}[2]z^{K-2} \ldots \\
&\quad + a_{\text{obs}}[k]z^{K-k} \ldots \\
&\quad + a_{\text{obs}}[K-2]z^2 + a_{\text{obs}}[K-1]z + a_{\text{obs}}[K]
\end{aligned} \tag{6.8a}$$

thus
$g_{\text{obs}}[0] = -a_{\text{obs}}[K-0]$
$g_{\text{obs}}[1] = -a_{\text{obs}}[K-1]$
$g_{\text{obs}}[2] = -a_{\text{obs}}[K-2]$
$\vdots$
$g_{\text{obs}}[k] = -a_{\text{obs}}[K-k]$
$\vdots$
$g_{\text{obs}}[K-2] = -a_{\text{obs}}[2]$
$g_{\text{obs}}[K-1] = -a_{\text{obs}}[1].$ \hfill (6.8b)

The elements of $\boldsymbol{g}_{\text{prc}}$ are obtained from the poles of the process model as follows:

$$\begin{aligned}
\mathcal{A}_{\text{prc}}(z) &= \prod_{k=0}^{K-1}(z - \rho_k) \\
&= z^K + a_{\text{prc}}[1]z^{K-1} + a_{\text{prc}}[2]z^{K-2} \ldots \\
&\quad + a_{\text{prc}}[k]z^{K-k} \ldots \\
&\quad + a_{\text{prc}}[K-2]z^2 + a_{\text{prc}}[K-1]z + a_{\text{prc}}[K]
\end{aligned} \tag{6.9a}$$

thus
$g_{\text{prc}}[0] = -a_{\text{prc}}[K-0]$
$g_{\text{prc}}[1] = -a_{\text{prc}}[K-1]$
$g_{\text{prc}}[2] = -a_{\text{prc}}[K-2]$
$\vdots$
$g_{\text{prc}}[k] = -a_{\text{prc}}[K-k]$
$\vdots$
$g_{\text{prc}}[K-2] = -a_{\text{prc}}[2]$
$g_{\text{prc}}[K-1] = -a_{\text{prc}}[1].$ \hfill (6.9b)

With $\mathcal{K}^{\text{pcf}}$ determined in canonical coordinates, it is transformed back into kinematic coordinates using

$$\mathcal{K}^{\text{kin}} = \mathbb{T}^{\text{kin}\leftarrow\text{pcf}}_{\text{prc}} \mathcal{K}^{\text{pcf}}. \tag{6.10}$$

It has so far been assumed that the $\mathbb{T}_{\text{prc}}^{\text{pcf}\leftarrow\text{kin}}$ transform and its inverse $\mathbb{T}_{\text{prc}}^{\text{kin}\leftarrow\text{pcf}} = \left\{\mathbb{T}_{\text{prc}}^{\text{pcf}\leftarrow\text{kin}}\right\}^{-1}$ are known. It is clearly possible to compute $\mathcal{K}^{\text{pcf}}$ without this knowledge; however, it is not possible to realize the observer, because the $C_{\text{obs}}^{\text{pcf}}$ operator for this coordinate system is unknown. Fortunately, the required transforms are readily found using

$$\mathbb{T}_{\text{prc}}^{\text{kin}\leftarrow\text{pcf}} = \{\mathcal{O}_{\text{prc}}^{\text{kin}}\}^{-1}\mathcal{O}_{\text{prc}}^{\text{pcf}} \tag{6.11}$$

where $\mathcal{O}$ is the $K \times K$ observability matrix for the $\langle C, G \rangle$ pair, with the $k$th row equal to $CG^k$ (for $0 \leq k < K$). The observability matrices for the kinematic and PCF systems are constructed using the $\langle C_{\text{prd}}^{\text{kin}}, G_{\text{prc}}^{\text{kin}} \rangle$ and $\langle C_{\text{prd}}^{\text{pcf}}, G_{\text{prc}}^{\text{pcf}} \rangle$ definitions provided above in (4.4b) & (3.2h) and (6.2b) & (6.2a), respectively. The transform only exists if $\langle C_{\text{prd}}^{\text{kin}}, G_{\text{prc}}^{\text{kin}} \rangle$ is observable, i.e. if $\text{rank}\{\mathcal{O}_{\text{prc}}^{\text{kin}}\} = K$.

The availability of the Ackermann or the Bass-Gura formulae [10],[11], which permit the direct computation of the gain vector, means that the coordinate transformation and working described in this section is not essential for observer design. However, an appreciation of this process is useful because it leads to: the various coordinate-system options for single-input/multiple-output (SIMO) observers and SISO filters, which are discussed in the next section; the analysis procedures for SISO filters, discussed in Section 7; and the various realization options for SISO filters, discussed in Section 8. Note that the term 'observer' is used here to refer to an estimator of the full kinematic state vector, i.e. SIMO case. Whereas, the term 'filter' (without the 'Kalman' qualifier) is used here to refer to an estimator of a single element of the kinematic state vector, e.g. position (smoother) or velocity (differentiator), i.e. the SISO case.

# 7. Observer realization

The observer may now be realized using any of the following:

1) with $\langle C_{\text{obs}}^{\text{kin}}, G_{\text{obs}}^{\text{kin}}, H_{\text{obs}}^{\text{kin}} \rangle$, directly in kinematic coordinates (**KIN**)

2) with $\langle C_{\text{obs}}^{\text{pcf}}, G_{\text{obs}}^{\text{pcf}}, H_{\text{obs}}^{\text{pcf}} \rangle$, in a coordinate system for a process canonical form (**PCF**), w.r.t. the $\langle C_{\text{prd}}^{\text{kin}}, G_{\text{prc}}^{\text{kin}} \rangle$ pair, with the kinematic states extracted from the internal filter states using $\mathbb{T}_{\text{prc}}^{\text{kin} \leftarrow \text{pcf}}$

3) with $\langle C_{\text{obs}}^{\text{ocf}}, G_{\text{obs}}^{\text{ocf}}, H_{\text{obs}}^{\text{ocf}} \rangle$ in a new coordinate system for an observable canonical form (**OCF**), w.r.t the $\langle C_{\text{obs}}^{\text{kin}}, G_{\text{obs}}^{\text{kin}} \rangle$ pair, with the kinematic states extracted from the internal filter states using $\mathbb{T}_{\text{obs}}^{\text{kin} \leftarrow \text{ocf}}$ or

4) with $\langle C_{\text{obs}}^{\text{ccf}}, G_{\text{obs}}^{\text{ccf}}, H_{\text{obs}}^{\text{ccf}} \rangle$, in a new coordinate system for a controllable canonical form (**CCF**), w.r.t the $\langle G_{\text{obs}}^{\text{kin}}, H_{\text{obs}}^{\text{kin}} \rangle$ pair, with the kinematic states extracted from the internal filter states using $\mathbb{T}_{\text{obs}}^{\text{kin} \leftarrow \text{ccf}}$.

Other possibilities such as the diagonal canonical form and block-diagonal forms are not considered here.

A generic block-diagram for the observer, which is fully reduced using (4.6) and independent of the coordinate system, is given in Figure 5. This representation may be used to realize the observer using the appropriate $\langle C_{\text{obs}}, G_{\text{obs}}, H_{\text{obs}} \rangle$ triplet and corresponding $\mathbb{T}_{\text{obs}}$, for the chosen coordinate system. Reduced computational complexity is the main reason for using one of the canonical forms; however, due to the non-diagonal $\mathbb{T}_{\text{obs}}$ matrix for these non-kinematic systems, this is only the case for SISO systems (i.e. filters), where the full kinematic state estimate $\widehat{w}_{\text{prc}}[n]$, is not required for all $n$.

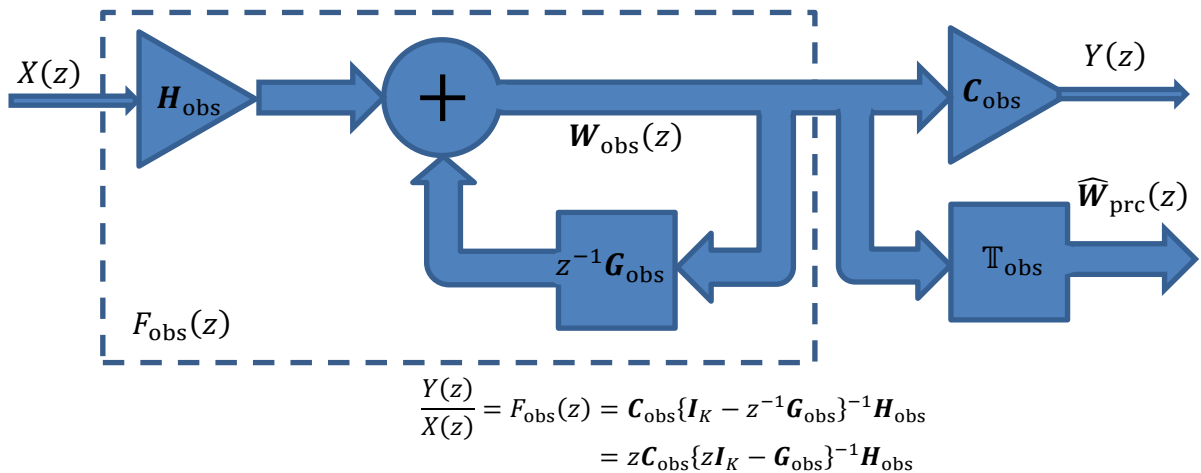

*Figure 5 - Reduced block diagram for the generic (Z-transformed) observer system. The internal coordinates of the observer are unspecified, as they depend on the realization. For kinematic (KIN) coordinates: $\mathbb{T}_{obs} = I_K$. For PCF w.r.t $\langle C_{prd}^{kin}, G_{prc}^{kin} \rangle$: $\mathbb{T}_{obs} = \mathbb{T}_{prc}^{kin \leftarrow pcf}$. For OCF w.r.t $\langle C_{obs}^{kin}, G_{obs}^{kin} \rangle$: $\mathbb{T}_{obs} = \mathbb{T}_{obs}^{kin \leftarrow ocf}$. For CCF w.r.t $\langle G_{obs}^{kin}, H_{obs}^{kin} \rangle$: $\mathbb{T}_{obs} = \mathbb{T}_{obs}^{kin \leftarrow ccf}$. The corresponding inverse transforms are required to initialize the observer.*

## 7.1 Kinematic form (KIN)

With $\mathbb{T}_{\text{prc}}^{\text{kin}\leftarrow\text{pcf}}$ determined – see (6.11) – it may be used in (6.10) to transform the gain vector back into kinematic coordinates, for use in the recursion of (4.6), i.e.

$$w_{\text{obs}}^{\text{kin}}[n] = G_{\text{obs}}^{\text{kin}} w_{\text{obs}}^{\text{kin}}[n-1] + H_{\text{obs}}^{\text{kin}} x[n] \text{ and} \tag{7.1.1a}$$
$$y[n] = C_{\text{obs}}^{\text{kin}} w_{\text{obs}}^{\text{kin}}[n] \text{ where} \tag{7.1.1b}$$
$$G_{\text{obs}}^{\text{kin}} = G_{\text{prc}}^{\text{kin}} - \mathcal{K}^{\text{kin}} C_{\text{prd}}^{\text{kin}}, \quad H_{\text{obs}}^{\text{kin}} = \mathcal{K}^{\text{kin}} \tag{7.1.1c}$$
$$C_{\text{obs}}^{\text{kin}} = C_{\text{prc}}^{\text{kin}} \{G_{\text{prc}}^{\text{kin}}\}^{-q} \text{ and} \tag{7.1.1d}$$
$$w_{\text{obs}}^{\text{kin}}[0] = \{C_{\text{prc}}^{\text{kin}}\}^T x[0] \text{ for initialization.} \tag{7.1.1e}$$

For $K = 2$, with both poles at $p$, the LSS matrices of the $\alpha - \beta$ filter are computed using

$$\mathcal{K}^{\text{kin}} = \begin{bmatrix} \alpha \\ \beta/T_s \end{bmatrix} \text{ where} \tag{7.1.2a}$$
$$\alpha = 1 - p^2 \text{ and} \tag{7.1.2b}$$
$$\beta = (p-1)^2 \text{ with} \tag{7.1.2c}$$
$$G_{\text{prc}}^{\text{kin}} = \begin{bmatrix} 1 & T_s \\ 0 & 1 \end{bmatrix} \text{ and} \tag{7.1.2d}$$
$$C_{\text{prc}}^{\text{kin}} = \begin{bmatrix} 1 & 0 \end{bmatrix}. \tag{7.1.2e}$$

For $K = 3$, with all poles at $p$, the LSS matrices of the $\alpha - \beta - \gamma$ filter are computed using

$$\mathcal{K}^{\text{kin}} = \begin{bmatrix} \alpha \\ \beta/T_s \\ \gamma/2T_s^2 \end{bmatrix} \text{ where} \tag{7.1.3a}$$
$$\alpha = 1 - p^3 \tag{7.1.3b}$$
$$\beta = \frac{3}{2}(p-1)^2(p+1) \text{ and} \tag{7.1.3c}$$
$$\beta = -2(p-1)^3 \text{ with} \tag{7.1.3d}$$
$$G_{\text{prc}}^{\text{kin}} = \begin{bmatrix} 1 & T_s & T_s^2/2 \\ 0 & 1 & T_s \\ 0 & 0 & 1 \end{bmatrix} \text{ and} \tag{7.1.3e}$$
$$C_{\text{prc}}^{\text{kin}} = \begin{bmatrix} 1 & 0 & 0 \end{bmatrix}. \tag{7.1.3f}$$

We now have everything we need to realize the observer and the disinterested reader need not proceed further.

This coordinate system is convenient for offline process-modelling and online signal-analysis purposes because it maintains the direct one-to-one mapping of the discrete-time states used in (3.3) to the continuous-time states used in (3.1). Thus, the kinematic coordinates have physical significance and a linear operation is not required to transform the internal states of the filter. However, the simpler (but fundamentally equivalent) structure of the state-space equations resulting from the utilization of a canonical coordinate-system is: more convenient for the offline analysis of the observer's response; and more efficient for online filter implementation/realization because it reduces the number of arithmetic operations through the use of sparse operators; furthermore, it allows standard/generic interfaces to hardware-optimized filtering libraries to be used, with the internal state vector of the filter interpreted as simply being a sequence of delay registers.

## 7.2  Process canonical form (PCF)

The transform and the LSS observer in this coordinate system were required in Section 6 to derive the gain vector that places the observer poles for the desired convergence behavior. The procedure and the result are summarized here, mainly to establish the need for other canonical forms.

The $\mathbb{T}_{\text{prc}}^{\text{pcf}\leftarrow\text{kin}}$ operator changes the kinematic coordinate system in a way that transforms the observable $\langle C_{\text{prd}}^{\text{kin}}, G_{\text{prc}}^{\text{kin}}\rangle$ pair into a canonical $\langle C_{\text{prd}}^{\text{pcf}} G_{\text{prc}}^{\text{pcf}}\rangle$ pair; the $\mathbb{T}_{\text{prc}}^{\text{kin}\leftarrow\text{pcf}}$ operator reverses the transformation, i.e.

$$w_{\text{obs}}^{\text{pcf}} = \mathbb{T}_{\text{prc}}^{\text{pcf}\leftarrow\text{kin}} w_{\text{prc}}^{\text{kin}} \text{ and} \qquad (7.2.1a)$$

$$w_{\text{prc}}^{\text{kin}} = \mathbb{T}_{\text{prc}}^{\text{kin}\leftarrow\text{pcf}} w_{\text{obs}}^{\text{pcf}} \text{ such that} \qquad (7.2.1b)$$

$$\langle C_{\text{prd}}^{\text{pcf}} G_{\text{prc}}^{\text{pcf}}\rangle \xrightleftharpoons[\mathbb{T}_{\text{prc}}^{\text{kin}\leftarrow\text{pcf}}]{\mathbb{T}_{\text{prc}}^{\text{pcf}\leftarrow\text{kin}}} \langle C_{\text{prd}}^{\text{kin}}, G_{\text{prc}}^{\text{kin}}\rangle \text{ where} \qquad (7.2.1c)$$

$$\mathbb{T}_{\text{prc}}^{\text{kin}\leftarrow\text{pcf}} = \{\mathcal{O}_{\text{prc}}^{\text{kin}}\}^{-1} \mathcal{O}_{\text{prc}}^{\text{pcf}} \text{ and} \qquad (7.2.1d)$$

$$\mathbb{T}_{\text{prc}}^{\text{pcf}\leftarrow\text{kin}} = \{\mathbb{T}_{\text{prc}}^{\text{kin}\leftarrow\text{pcf}}\}^{-1}. \qquad (7.2.1e)$$

The required transforms are computed using $\mathcal{O}_{\text{prc}}^{\text{kin}}$ and $\mathcal{O}_{\text{prc}}^{\text{pcf}}$ which are the $K \times K$ observability matrices for the observable $\langle C_{\text{prd}}^{\text{kin}}, G_{\text{prc}}^{\text{kin}}\rangle$ and $\langle C_{\text{prd}}^{\text{pcf}}, G_{\text{prc}}^{\text{pcf}}\rangle$ pairs. The transform only exists if $\langle C_{\text{prd}}^{\text{kin}}, G_{\text{prc}}^{\text{kin}}\rangle$ is observable, i.e. if $\text{rank}\{\mathcal{O}_{\text{prc}}^{\text{kin}}\} = K$.

In this canonical coordinate system, we have the following:

$$w_{\text{obs}}^{\text{pcf}}[n] = G_{\text{obs}}^{\text{pcf}} w_{\text{obs}}^{\text{pcf}}[n-1] + H_{\text{obs}}^{\text{pcf}} x[n] \qquad (7.2.2a)$$

$$y[n] = C_{\text{obs}}^{\text{pcf}} w_{\text{obs}}^{\text{pcf}}[n] \text{ where} \qquad (7.2.2b)$$

$$G_{\text{obs}}^{\text{pcf}} = \begin{bmatrix} \mathbf{0}_{1\times(K-1)} & \\ I_{(K-1)\times(K-1)} & g_{\text{obs}} \end{bmatrix}_{K\times K}, H_{\text{obs}}^{\text{pcf}} = \mathbb{T}_{\text{prc}}^{\text{pcf}\leftarrow\text{kin}} H_{\text{obs}}^{\text{kin}} = \mathcal{K}^{\text{pcf}} \qquad (7.2.2c)$$

$$C_{\text{obs}}^{\text{pcf}} = C_{\text{obs}}^{\text{kin}} \mathbb{T}_{\text{prc}}^{\text{kin}\leftarrow\text{pcf}} \text{ and} \qquad (7.2.2d)$$

$$w_{\text{obs}}^{\text{pcf}}[0] = \mathbb{T}_{\text{prc}}^{\text{pcf}\leftarrow\text{kin}} w_{\text{obs}}^{\text{kin}}[0] \text{ for state initialization and} \qquad (7.2.2e)$$

$$w_{\text{obs}}^{\text{kin}}[n] = \mathbb{T}_{\text{prc}}^{\text{kin}\leftarrow\text{pcf}} w_{\text{obs}}^{\text{pcf}}[n] \text{ for state extraction.} \qquad (7.2.2f)$$

The correctness of the transform may be confirmed using:

$$G_{\text{prc}}^{\text{pcf}} = \mathbb{T}_{\text{prc}}^{\text{pcf}\leftarrow\text{kin}} G_{\text{prc}}^{\text{kin}} \mathbb{T}_{\text{prc}}^{\text{kin}\leftarrow\text{pcf}} \text{ and} \qquad (7.2.3a)$$

$$C_{\text{prd}}^{\text{pcf}} = C_{\text{prd}}^{\text{kin}} \mathbb{T}_{\text{prc}}^{\text{kin}\leftarrow\text{pcf}}. \qquad (7.2.3b)$$

Note that in this coordinate system, the process pair $\langle C_{\text{prd}}^{\text{kin}}, G_{\text{prc}}^{\text{kin}} \rangle$ has a canonical form; however, the observer system does not: because $G_{\text{obs}}^{\text{pcf}}$ has been simplified but both $H_{\text{obs}}^{\text{pcf}}$ and $C_{\text{obs}}^{\text{pcf}}$ are non-trivial. We therefore have a non-canonical coordinates-system that does not minimize the complexity associated with each update in (7.2.2a) & (7.2.2b) *and* a non-kinematic coordinate-system with the extra complexity associated with state extraction in (7.2.2f). Thus, PCF is useful for observer design, but it is not recommended for observer realization.

Process canonical form (PCF) is a non-standard term that is used here to distinguish it from the standard observer-canonical form (OCF) that is discussed in the next subsection. Both forms end in a canonical $\langle C, G \rangle$ pair with the same structure; however, PCF and OCF begin with the consideration of the $\langle C_{\text{prd}}^{\text{kin}}, G_{\text{prc}}^{\text{kin}} \rangle$ and $\langle C_{\text{obs}}^{\text{kin}}, G_{\text{obs}}^{\text{kin}} \rangle$ pairs, respectively.

## 7.3 Observable canonical form (OCF)

The coordinate transform applied in Section 5 was used to expose the coefficients of the $\mathcal{A}_{\text{prc}}(z)$ polynomial in the $G_{\text{prc}}$ matrix and to focus the action of the $C_{\text{prd}}$ operator so that the gain vector $\mathcal{K}$ only acts on the desired column of $G_{\text{prc}}$. A similar transformation is applied here; however, this is now done to simplify the $C_{\text{obs}}$ and $G_{\text{obs}}$ operators for online computational efficiency and to expose the coefficients of the $\mathcal{B}_{\text{obs}}(z)$ polynomial in the $H_{\text{obs}}$ operator for offline response analysis (see Section 8).

As discussed in the previous subsections, the observer equations in (4.6a) & (4.6b) are not in OCF after the gain vector is introduced to form $G_{\text{obs}}^{\text{kin}}$ and when the $q$-parameterized $C_{\text{obs}}^{\text{kin}}$ operator is used; therefore, a new transform $\mathbb{T}_{\text{obs}}^{\text{kin} \leftarrow \text{ocf}}$ is required. It is computed using a procedure that is analogous to the one that was used previously to compute $\mathbb{T}_{\text{prc}}^{\text{kin} \leftarrow \text{ocf}}$; with observability matrices for the observable $\langle G_{\text{obs}}^{\text{kin}}, H_{\text{obs}}^{\text{kin}} \rangle$ pair used instead of observability matrices for the observable $\langle C_{\text{prc}}^{\text{kin}}, G_{\text{prc}}^{\text{kin}} \rangle$ pair.

The $\mathbb{T}_{\text{obs}}^{\text{ocf} \leftarrow \text{kin}}$ operator changes the kinematic coordinate system in a way that transforms the observable $\langle C_{\text{obs}}^{\text{kin}}, G_{\text{obs}}^{\text{kin}} \rangle$ pair into a canonical $\langle C_{\text{obs}}^{\text{ocf}}, G_{\text{obs}}^{\text{ocf}} \rangle$ pair; the $\mathbb{T}_{\text{obs}}^{\text{kin} \leftarrow \text{ocf}}$ operator reverses the transformation, i.e.

$$w_{\text{obs}}^{\text{ocf}} = \mathbb{T}_{\text{obs}}^{\text{ocf} \leftarrow \text{kin}} w_{\text{obs}}^{\text{kin}} \text{ and} \tag{7.3.1a}$$

$$w_{\text{obs}}^{\text{kin}} = \mathbb{T}_{\text{obs}}^{\text{kin} \leftarrow \text{ocf}} w_{\text{obs}}^{\text{ocf}} \text{ such that} \tag{7.3.1b}$$

$$\langle C_{\text{obs}}^{\text{kin}}, G_{\text{obs}}^{\text{kin}} \rangle \underset{\mathbb{T}_{\text{obs}}^{\text{ocf} \leftarrow \text{kin}}}{\overset{\mathbb{T}_{\text{obs}}^{\text{kin} \leftarrow \text{ocf}}}{\rightleftarrows}} \langle C_{\text{obs}}^{\text{ocf}}, G_{\text{obs}}^{\text{ocf}} \rangle \text{ where} \tag{7.3.1c}$$

$$\mathbb{T}_{\text{obs}}^{\text{kin} \leftarrow \text{ocf}} = \{\mathcal{O}_{\text{obs}}^{\text{kin}}\}^{-1} \mathcal{O}_{\text{obs}}^{\text{ocf}} \tag{7.3.1d}$$

$$\mathbb{T}_{\text{obs}}^{\text{ocf} \leftarrow \text{kin}} = \{\mathbb{T}_{\text{obs}}^{\text{kin} \leftarrow \text{ocf}}\}^{-1}. \tag{7.3.1e}$$

The required transforms are computed using $\mathcal{O}_{\text{obs}}^{\text{kin}}$ and $\mathcal{O}_{\text{obs}}^{\text{ocf}}$ which are the $K \times K$ observability matrices for the observable $\langle C_{\text{obs}}^{\text{kin}}, G_{\text{obs}}^{\text{kin}} \rangle$ and $\langle C_{\text{obs}}^{\text{ocf}}, G_{\text{obs}}^{\text{ocf}} \rangle$ pairs. The transform only exists if $\langle C_{\text{obs}}^{\text{kin}}, G_{\text{obs}}^{\text{kin}} \rangle$ is observable, i.e. if $\text{rank}\{\mathcal{O}_{\text{obs}}^{\text{kin}}\} = K$.

In this canonical coordinate system, we have the following:

$$w_{\text{obs}}^{\text{ocf}}[n] = G_{\text{obs}}^{\text{ocf}} w_{\text{obs}}^{\text{ocf}}[n-1] + H_{\text{obs}}^{\text{ocf}} x[n] \tag{7.3.2a}$$

$$y[n] = C_{\text{obs}}^{\text{ocf}} w_{\text{obs}}^{\text{ocf}}[n] \text{ where} \tag{7.3.2b}$$

$$G_{\text{obs}}^{\text{ocf}} = \begin{bmatrix} \mathbf{0}_{1 \times (K-1)} & \\ I_{(K-1) \times (K-1)} & g_{\text{obs}} \end{bmatrix}_{K \times K}, \quad H_{\text{obs}}^{\text{ocf}} = \mathbb{T}_{\text{obs}}^{\text{ocf} \leftarrow \text{kin}} H_{\text{obs}}^{\text{kin}} \tag{7.3.2c}$$

$$C_{\text{obs}}^{\text{ocf}} = [\mathbf{0}_{1 \times (K-1)} \quad 1]_{1 \times K} \text{ (by definition)} \tag{7.3.2d}$$

$$w_{\text{obs}}^{\text{ocf}}[0] = \mathbb{T}_{\text{obs}}^{\text{ocf} \leftarrow \text{kin}} w_{\text{obs}}^{\text{kin}}[0] \text{ for state initialization and} \tag{7.3.2e}$$

$$w_{\text{obs}}^{\text{kin}}[n] = \mathbb{T}_{\text{obs}}^{\text{kin} \leftarrow \text{ocf}} w_{\text{obs}}^{\text{ocf}}[n] \text{ for state extraction.} \tag{7.3.2f}$$

The correctness of the transform may be confirmed using:

$$G_{\text{obs}}^{\text{ocf}} = \mathbb{T}_{\text{obs}}^{\text{ocf} \leftarrow \text{kin}} G_{\text{obs}}^{\text{kin}} \mathbb{T}_{\text{obs}}^{\text{kin} \leftarrow \text{ocf}} \text{ and} \tag{7.3.3a}$$

$$C_{\text{obs}}^{\text{ocf}} = C_{\text{obs}}^{\text{kin}} \mathbb{T}_{\text{obs}}^{\text{kin} \leftarrow \text{ocf}}. \tag{7.3.3b}$$

For $K = 2$, with both poles at $p$, the LSS matrices of the $\alpha - \beta$ filter are

$$H_{\text{obs}}^{\text{ocf}} = \begin{bmatrix} 2p(p-1) \\ 1 - p^2 \end{bmatrix} \tag{7.3.4a}$$

$$G_{\text{obs}}^{\text{ocf}} = \begin{bmatrix} 0 & -p^2 \\ 1 & 2p \end{bmatrix} \text{ and} \tag{7.3.4b}$$

$$C_{\text{obs}}^{\text{ocf}} = [0 \quad 1]. \tag{7.3.4c}$$

In this case, $p$ is the same as the $\theta$ parameter used in the fading-memory filters of second degree in [25] and the 'g-h filters' in [26].

For $K = 3$, with all poles at $p$, the LSS matrices of the $\alpha - \beta - \gamma$ filter are

$$H_{\text{obs}}^{\text{ocf}} = \begin{bmatrix} -3p^2(p-1) \\ 3p(p^2 - 1) \\ 1 - p^3 \end{bmatrix} \tag{7.3.5a}$$

$$G_{\text{obs}}^{\text{ocf}} = \begin{bmatrix} 0 & 0 & p^3 \\ 1 & 0 & -3p^2 \\ 0 & 1 & 3p \end{bmatrix} \text{ and} \tag{7.3.5b}$$

$$C_{\text{obs}}^{\text{ocf}} = [0 \quad 0 \quad 1]. \tag{7.3.5c}$$

## 7.4 Controllable canonical form (CCF)

This form is motivated by the same considerations that led to the OCF covered in the previous subsection and the result is similar. However, in the current open-loop context (i.e. state estimation without a plant to control) observability (i.e. the ability to passively estimate all internal states) is important but controllability (i.e. the ability to actively drive all internal states) is not. Therefore, the difference between OCF and CCF is not as significant as it might otherwise be. Thus, choosing between OCF and CCF for observer realization is somewhat arbitrary/academic and more to do with convention and preference than efficiency and performance. An observer that is realized in CCF will behave in the same way as an observer that is realized in OCF; however, because the internal coordinate system and structure are different, the internal states must be interpreted differently. This distinction is important for filter initialization and kinematic state extraction. This form is not essential for analysis or realization and it is presented here for completeness.

The $\mathbb{T}_{\text{obs}}^{\text{ccf}\leftarrow\text{kin}}$ operator changes the kinematic coordinate system in a way that transforms the controllable $\langle G_{\text{obs}}^{\text{kin}}, H_{\text{obs}}^{\text{kin}}\rangle$ pair into a canonical $\langle G_{\text{obs}}^{\text{ccf}}, H_{\text{obs}}^{\text{ccf}}\rangle$ pair; the $\mathbb{T}_{\text{obs}}^{\text{kin}\leftarrow\text{ccf}}$ operator reverses the transformation, i.e.

$$w_{\text{obs}}^{\text{ccf}} = \mathbb{T}_{\text{obs}}^{\text{ccf}\leftarrow\text{kin}} w_{\text{obs}}^{\text{kin}} \text{ and} \qquad (7.4.1a)$$
$$w_{\text{obs}}^{\text{kin}} = \mathbb{T}_{\text{obs}}^{\text{kin}\leftarrow\text{ccf}} w_{\text{obs}}^{\text{ccf}} \text{ such that} \qquad (7.4.1b)$$
$$\langle G_{\text{obs}}^{\text{kin}}, H_{\text{obs}}^{\text{kin}}\rangle \underset{\mathbb{T}_{\text{obs}}^{\text{ccf}\leftarrow\text{kin}}}{\overset{\mathbb{T}_{\text{obs}}^{\text{kin}\leftarrow\text{ccf}}}{\rightleftarrows}} \langle G_{\text{obs}}^{\text{ccf}}, H_{\text{obs}}^{\text{ccf}}\rangle \text{ where} \qquad (7.4.1c)$$
$$\mathbb{T}_{\text{obs}}^{\text{kin}\leftarrow\text{ccf}} = \mathcal{C}_{\text{obs}}^{\text{kin}}\{\mathcal{C}_{\text{obs}}^{\text{ccf}}\}^{-1} \text{ and} \qquad (7.4.1d)$$
$$\mathbb{T}_{\text{obs}}^{\text{ccf}\leftarrow\text{kin}} = \{\mathbb{T}_{\text{obs}}^{\text{kin}\leftarrow\text{ccf}}\}^{-1}. \qquad (7.4.1e)$$

The required transforms are computed using $\mathcal{C}_{\text{obs}}^{\text{kin}}$ and $\mathcal{C}_{\text{obs}}^{\text{ccf}}$ which are the $K \times K$ controllability matrices for the controllable $\langle G_{\text{obs}}^{\text{kin}}, H_{\text{obs}}^{\text{kin}}\rangle$ and $\langle G_{\text{obs}}^{\text{ccf}}, H_{\text{obs}}^{\text{ccf}}\rangle$ pairs, respectively. For a given $\langle G, H\rangle$ pair, the $k$th column of $\mathcal{C}$ is equal to $G^k H$ (for $0 \leq k < K$). The transform only exists if $\langle G_{\text{obs}}^{\text{kin}}, H_{\text{obs}}^{\text{kin}}\rangle$ is controllable, i.e. if $\text{rank}\{\mathcal{C}_{\text{obs}}^{\text{kin}}\} = K$.

In this canonical coordinate system, we have the following:

$$w_{\text{obs}}^{\text{ccf}}[n] = G_{\text{obs}}^{\text{ccf}} w_{\text{obs}}^{\text{ccf}}[n-1] + H_{\text{obs}}^{\text{ccf}} x[n] \qquad (7.4.2a)$$
$$y[n] = C_{\text{obs}}^{\text{ccf}} w_{\text{obs}}^{\text{ccf}}[n] \text{ where} \qquad (7.4.2b)$$
$$G_{\text{obs}}^{\text{ccf}} = \begin{bmatrix} g_{\text{obs}} \\ I_{(K-1)\times(K-1)} & 0_{(K-1)\times 1}\end{bmatrix}_{K\times K}, H_{\text{obs}}^{\text{ccf}} = \begin{bmatrix}1 \\ 0_{(K-1)\times 1}\end{bmatrix}_{K\times 1} \text{ (by definition)} \qquad (7.4.2c)$$
$$C_{\text{obs}}^{\text{ccf}} = C_{\text{obs}}^{\text{kin}} \mathbb{T}_{\text{obs}}^{\text{kin}\leftarrow\text{ccf}} \qquad (7.4.2d)$$
$$w_{\text{obs}}^{\text{ccf}}[0] = \mathbb{T}_{\text{obs}}^{\text{ccf}\leftarrow\text{kin}} w_{\text{obs}}^{\text{kin}}[0] \text{ for state initialization} \qquad (7.4.2e)$$
$$w_{\text{obs}}^{\text{kin}}[n] = \mathbb{T}_{\text{obs}}^{\text{kin}\leftarrow\text{ccf}} w_{\text{obs}}^{\text{ccf}}[n] \text{ for state extraction.} \qquad (7.4.2f)$$

The correctness of the transform may be confirmed using:

$$G_{\text{obs}}^{\text{ccf}} = \mathbb{T}_{\text{obs}}^{\text{ccf}\leftarrow\text{kin}} G_{\text{obs}}^{\text{kin}} \mathbb{T}_{\text{obs}}^{\text{kin}\leftarrow\text{ccf}} \text{ and} \qquad (7.4.3a)$$
$$H_{\text{obs}}^{\text{ccf}} = \mathbb{T}_{\text{obs}}^{\text{ccf}\leftarrow\text{kin}} H_{\text{obs}}^{\text{kin}}. \qquad (7.4.3b)$$

SISO smoothers (or differentiators), reached via either of the canonical forms (i.e. OCF or CCF), are worth considering if the internal states are not required for each new sample. In such cases, optimized filtering routines may be used, e.g. MATLAB's `y = filter(b,a,x)` function.

# 8. Filter analysis

This section applies to any discrete-time LSS representation of a system, therefore there is no need to use the 'prc' and 'obs' and subscripts here. These were only required to distinguish between the natural process and the synthetic observer for the gain vector derivation.

As elucidated in [15], $z$-plane analysis is not a perverse theoretical abstraction; rather, it is a mathematical construct of profound practical importance, as $\mathcal{H}(z)$ embodies all aspects of the filter response; namely, the (steady-state) frequency response $H(\omega)$, the (transient) impulse response $h[n]$, and the linear difference equation (LDE) of the time-domain realization. Furthermore, $\mathcal{Z}$-transformed transfer-functions permit the analysis of stability/instability and the algebraic manipulation of delay operations so that convolution becomes (polynomial) multiplication, which is utilized in the rearrangement and reduction of block-diagram representations (e.g. Figure 4 to Figure 5) for the analysis of a system-of-systems. These elements are brought together and discussed in this section.

## 8.1 Identification of system zeros

The discrete-time (rational) transfer-functions $\mathcal{H}(z)$, of LSS systems defined using (4.7) or (4.8) have a numerator polynomial $\mathcal{B}(z)$, with an order that is less than or equal to the order of $\mathcal{A}(z)$, i.e. they are 'proper'. Furthermore when (4.7) is used, which is the form preferred in this tutorial, all powers of $z$ in $\mathcal{B}(z)$ are greater than zero, so that $\boldsymbol{D}$ in (4.7b) is always zero. In this case, the coefficients of the $\mathcal{B}(z)$ and $\mathcal{A}(z)$ polynomials, which define the zeros and poles of the filter respectively, are extracted from the discrete-time LSS representation of the system in OCF or CCF, as follows:

$$\boldsymbol{G}^{\text{ocf}} = \begin{bmatrix} 0 & 0 & \cdots & 0 & 0 & -a[K-0] \\ 1 & 0 & \cdots & 0 & 0 & -a[K-1] \\ 0 & 1 & \cdots & 0 & 0 & -a[K-2] \\ \vdots & \vdots & \ddots & \vdots & \vdots & \vdots \\ 0 & 0 & \cdots & 1 & 0 & -a[2] \\ 0 & 0 & \cdots & 0 & 1 & -a[1] \end{bmatrix}_{K \times K}, \boldsymbol{H}^{\text{ocf}} = \begin{bmatrix} b[K-1] \\ b[K-2] \\ b[K-3] \\ \vdots \\ b[1] \\ b[0] \end{bmatrix}_{K \times 1}$$

$$\boldsymbol{C}^{\text{ocf}} = [0 \quad 0 \quad 0 \quad \cdots \quad 0 \quad 1]_{1 \times K};  \tag{8.1.1}$$

$$\boldsymbol{G}^{\text{ccf}} = \begin{bmatrix} -a[1] & -a[2] & \cdots & -a[K-2] & -a[K-1] & -a[K-0] \\ 1 & 0 & \cdots & 0 & 0 & 0 \\ 0 & 1 & \cdots & 0 & 0 & 0 \\ \vdots & \vdots & \ddots & \vdots & \vdots & \vdots \\ 0 & 0 & \cdots & 1 & 0 & 0 \\ 0 & 0 & \cdots & 0 & 1 & 0 \end{bmatrix}_{K \times K}, \boldsymbol{H}^{\text{ccf}} = \begin{bmatrix} 1 \\ 0 \\ \vdots \\ 0 \\ 0 \\ 0 \end{bmatrix}_{K \times 1}$$

$$\boldsymbol{C}^{\text{ccf}} = [\ b[0] \quad b[1] \quad \cdots \quad b[K-3] \quad b[K-2] \quad b[K-1]]_{1 \times K}. \tag{8.1.2}$$

As exploited in Section 5, the $\boldsymbol{a}$ coefficients of $\mathcal{A}(z)$ are arranged along the last column of $\boldsymbol{G}^{\text{ocf}}$ (or first row of $\boldsymbol{G}^{\text{ccf}}$). Equations (8.1.1) & (8.1.2) show that the $\boldsymbol{b}$ coefficients of $\mathcal{B}(z)$ are similarly arranged in $\boldsymbol{H}^{\text{ocf}}$ or $\boldsymbol{C}^{\text{ccf}}$. Note that $\boldsymbol{b}$ is a vector of length $K+1$, indexed as $0 \leq k \leq$ $K$, containing the coefficients $b[k]$, with $b[K] = 0$. Note that the first and last coefficients of ***a*** & ***b*** do not appear above because they are required to be unity and zero, respectively, for the systems considered here (i.e. $a[0] = 1$ and $b[K] = 0$). Enforcement of the latter requirement for the systems considered here, disentangles the ***a*** & ***b*** coefficients, which yields expressions that are simpler than those found in standard texts.

It is important to appreciate that in these canonical forms, the elements of the state vector have no obvious connection to the corresponding elements of the state vector in the 'native' kinematic (KIN) coordinate system, which are simply derivatives of the position coordinate w.r.t time, reached by discretizing the continuous-time LSS representation. Canonical coordinates are essential for design/analysis and they are optional for realization. In canonical coordinates, the filter states are simply an internal mechanism (i.e. delay 'registers') for the application of poles and zeros in a discrete-time system. Physical (e.g. Kinematic) coordinates are used to determine where those poles and zeros should be to achieve the desired objective. As discussed at length above, linear transforms are applied to move from one representation to another. Unfortunately, there is no consistent convention in use for the various canonical forms between and within the (discrete-time and continuous-time) control, signal processing, and electronics, fields.

Discrete-time representations are unavoidable in digital systems and in these cases, delay indexing is the most obvious of interpretation of the canonical state vector. Unfortunately, LSS representations are more commonly used in control, where a mix of continuous-time and discrete-time systems are usually considered, which confuses the situation. The state-transition matrix (***G***) of canonical forms for delay-indexed discrete-time systems have a band of ones immediately *below* the diagonal, which shifts the elements *down* one place in the state vector on each timestep. In CCF, the input is stored in the 0th element of ***w*** when it initially enters the system and the system zeros are applied as the output is 'collected' over all elements in the state vector when it leaves the system. In OCF, the system zeros are applied when the input is 'dispersed' over all elements of the state vector as a new sample enters the system and the output is taken from the $(K-1)$th element of ***w*** when it leaves the system. Canonical forms with advance indexing have a band of ones immediately *above* the diagonal, which shifts the elements *up* one place in the state vector on each timestep. These forms are less intuitive thus they are not considered here. They are more likely to be found in texts on control and analogue electronics, where LSS representations of continuous-time systems are usually the focus. Curiously, this format is also used inside MATLAB's `filter()` function, presumably to maintain consistency across continuous-time and discrete-time LSS representations, thus care is required when internal filter states are accessed (i.e. during set and get operations).

Delay indexing is more intuitive for digital filter *implementation*; however as utilized below, advance indexing is more intuitive for discrete-time system *analysis* (in a non-causal form) because the elements of the state vector correspond to positive powers of $z$, i.e. the $k$th element of ***w*** is initially reached by applying an advance of $z^k$ to the input. The prevalence of advance indexing (i.e. positive powers of $z$) in the discrete control literature is probably used to be consistent with the derivative states used in continuous LSS formulations (i.e. positive powers of $s$). In both discrete-time and continuous-time systems, positive powers (i.e. sequences of advance and differentiation operations, respectively) *may* be used for system analysis but negative powers *must* be used for system realization (i.e. sequences of delay and integration operations, respectively).

The availability of the **b** & **a** coefficients permits the $\mathcal{B}(z)$ and $\mathcal{A}(z)$ polynomials to be defined, thus the transfer function $\mathcal{H}(z) = \mathcal{B}(z)/\mathcal{A}(z)$ may now be simply specified using

$$\mathcal{H}(z) = \frac{\mathcal{B}(z)}{\mathcal{A}(z)} = \frac{\sum_{k=0}^{K-1} b[k]z^{K-k}}{z^K + \sum_{k=1}^{K} a[k]z^{K-k}} \tag{8.1.3}$$

instead of the more complicated form in (5.4c). This non-causal form is more amenable to analysis because its zeros and poles are readily identified, by finding the roots of $\mathcal{B}(z)$ and $\mathcal{A}(z)$, respectively. Note that $\mathcal{A}(z)$ is already known because its roots (i.e. the observer poles) were specified at the outset of the design process; of however, $\mathcal{B}(z)$ (i.e. the observer zeros) have so-far been unknown. For our smoother/observer the poles are all equal to $p$ and the zeros are a function of both $p$ & $q$.

Dividing $\mathcal{A}(z)$ and $\mathcal{B}(z)$ by $z^K$ yields the causal equivalent of $\mathcal{H}(z)$, which is expressed using only delays

$$\mathcal{H}(z) = \frac{Y(z)}{X(z)} = \frac{\sum_{k=0}^{K-1} b[k]z^{-k}}{1 + \sum_{k=1}^{K} a[k]z^{-k}}. \tag{8.1.4}$$

(Note that the indexing convention used here for **b** & **a** assumes that the $k$th coefficient is for the $k$th power of $1/z$.)

The frequency response $H(\omega)$, of $\mathcal{H}(z)$ is computed in the usual way [15], with

$$H(\omega) = \mathcal{H}(z)|_{z=e^{i\omega}} = \frac{\sum_{k=0}^{K-1} b[k]e^{-i\omega k}}{1 + \sum_{k=1}^{K} a[k]e^{-i\omega k}} \tag{8.1.5}$$

where $\omega$ is the angular frequency (radians per sample) and $i$ is the imaginary unit, $i = \sqrt{-1}$. The frequency response reveals how the filter responds to a sinusoidal input of infinite extent, at steady state. As the filter is an LTI system, the output only contains components oscillating at the frequencies of components comprising the input, with each component multiplied by a complex scaling factor, for a phase and magnitude shift, as indicated by evaluating $H(\omega)$ at those frequencies. The frequency response is particularly useful for analyzing a tracking filter's response to a constant g (circular) manoeuvre in two Cartesian coordinates [4],[16].

## 8.2  Linear difference equation (LDE)

Matrix/vector operations do not need to be used to realize $\mathcal{H}(z)$. As $\boldsymbol{G}^{\text{ccf}}$ is now quite sparse, it may be implemented more efficiently in the time domain by rearranging (8.1.4) to yield

$$Y(z) = X(z) \sum_{k=0}^{K-1} b[k]z^{-k} - Y(z) \sum_{k=1}^{K} a[k]z^{-k}. \tag{8.2.1}$$

Application of the inverse $\mathcal{Z}$-transform ($n \leftarrow z$) then yields the linear-difference equation (LDE) for a SISO embodiment with an infinite-impulse-response (IIR) or finite-impulse-response (FIR) [15]

$$y(n) = \sum_{k=0}^{K-1} b[k]x[n-k] - \sum_{k=1}^{K} a[k]y[n-k]. \tag{8.2.2}$$

In the non-recursive FIR case, $a[k] = 0$ for $1 \leq k \leq K$.

In this form, the transient and steady-state response of the observer system may be analyzed using standard software tools, or by simply processing an input signal with the desired form (e.g. an impulse, step or ramp) and monitoring the response prior to steady state. The transient response of a second-order smoother, for various combinations of $p$ & $q$ is shown in Figure 6. For this simple ($\alpha - \beta$) smoother, it is possible to derive closed-form expressions for the LDE coefficients in (8.1.7). The expressions provided below were derived using weighted least-squares regression and the orthogonal discrete Laguerre polynomial [19]; however, the result is identical to the smoother derived using an observer.

$$b[0] = (qp + p - q + 1)(1 - p)$$
$$b[1] = -(qp + 2p - q)(1 - p)$$
$$b[2] = 0 \tag{8.2.3a}$$

$$a[0] = 1$$
$$a[1] = -2p$$
$$a[2] = p^2. \tag{8.2.3b}$$

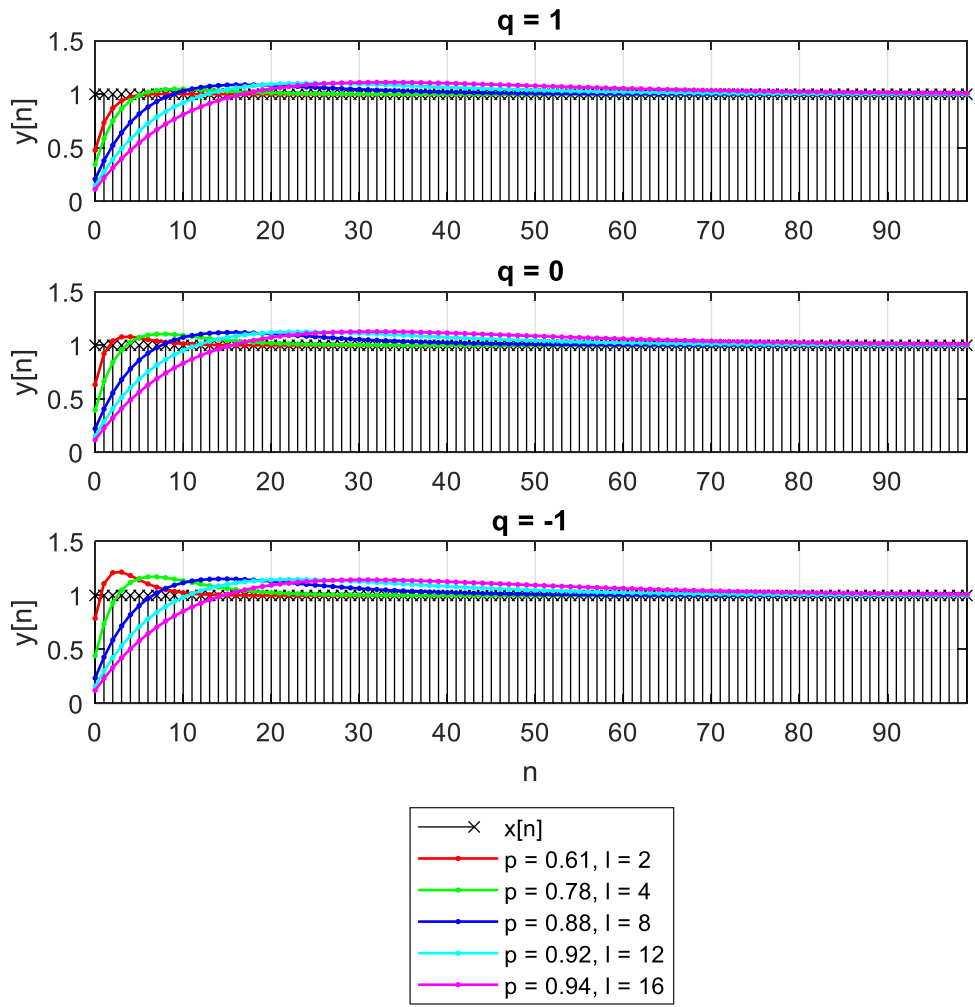

*Figure 6 – Unit step-response of various second order (K = 2) smoothers. The observers are parameterized using delays (q) of 1, 0 and -1, samples (top to bottom) and five different pole locations (p). The poles are set using $p = e^{-1/l}$ where l is the observer memory, in samples (see legend for values).*

## 8.3 Steady-state output

The output at steady state, of a stable filter, may also be evaluated analytically using the final-value theorem

$$\lim_{n \to \infty} y[n] = \frac{z-1}{z} Y(z)\Big|_{z=1} \text{ with} \tag{8.3.1a}$$
$$Y(z) = \mathcal{H}(z) X(z) \tag{8.3.1b}$$

(assuming the limit exists) where: $Y(z)$ is the $\mathcal{Z}$ transform of the output signal, $X(z)$ is the $\mathcal{Z}$ transform of the input signal, e.g. $X(z) = \frac{z}{z-1}$ and $X(z) = \frac{z}{(z-1)^2}$ for a unit step function and a unit ramp input (respectively) applied at $n = 0$, and $\mathcal{H}(z) = Y(z)/X(z)$, by definition. For a state estimator with $q = 0$ (i.e. zero lag/latency) and $K = 2$ (i.e. position and velocity states), it is essential to have a zero steady-state error for both step and ramp inputs. Given the use of the integrating process model and the absence of a plant in this observer-only state-estimation problem, this requirement is easily met when the observer model is matched to the process. Only the smoothers with shorter memories have time to reach steady state in Figure 6.

## 8.4 White-noise gain (WNG)

For a system with transfer function $\mathcal{H}(z)$, the white-noise gain (WNG) may be computed using its frequency response $H(\omega)$, or more conveniently, as a consequence of Parseval's theorem, using its impulse response $h[n]$, via a loop until convergence using

$$\text{WNG} = \frac{1}{2\pi} \int_{-\pi}^{+\pi} |H(\omega)|^2 d\omega = \sum_{n=0}^{\infty} |h[n]|^2. \tag{8.4.1}$$

The WNG indicates the extent to which white noise is attenuated (WNG < 1) or amplified (WNG > 1), at steady state, as it passes through the system, i.e.

$$\sigma_y^2 = \text{WNG} \cdot \sigma_x^2 \tag{8.4.2}$$

where $\sigma_x^2$ is the variance of the input noise (i.e. additive measurement error, $\sigma_R^2$) and $\sigma_y^2$ is the expected value of the squared estimation error (adjusted for an integer lag), i.e.

$$\sigma_x^2 = E\langle e^2 \rangle \text{ and} \tag{8.4.3a}$$
$$\sigma_y^2 = E\langle \varepsilon^2 \rangle \text{ with} \tag{8.4.3b}$$
$$\varepsilon = y_{\text{obs}}[n] - y_{\text{prc}}[n-q] = \hat{y}_{\text{prc}}[n-q] - y_{\text{prc}}[n-q]. \tag{8.4.3c}$$

The white noise $e[n]$ in Figure 3 is uncorrelated (by definition) and assumed to be zero-mean with finite variance but it need not be Gaussian for the Luenberger observers/smoothers considered here.

It is clearly desirable to have a low WNG, for good noise attenuation (i.e. low variance) at steady state; however, (8.4.1) indicates that this is difficult to achieve with a wide bandwidth, for a good transient response. Application of a fixed lag ($q > 0$) of a few samples generally decreases the WNG at the expense of system latency, provided the lag can be supported by the memory of the observer (using $q < l$ is recommended). Application of a phase lead ($q < 0$) is useful for anticipating future events and planning appropriate actions; however, this is achieved at the expense of an increases the WNG. A long prediction horizon with a short filter memory amplifies noise (WNG > 1). The WNG of the smoothers in Figure 6 is provided in Table 1.

*Table 1 – White-noise gain (WNG) for the smoothers in Figure 7 and Figure 8. WNG decreases: as the observer memory (l) increases, as the observer poles (p) approach unity, and as the lag (q) increases.*

|        | $l = 2$      | $l = 4$      | $l = 8$      | $l = 12$     | $l = 16$     |
|--------|--------------|--------------|--------------|--------------|--------------|
|        | $p = 0.6065$ | $p = 0.7788$ | $p = 0.8825$ | $p = 0.9200$ | $p = 0.9394$ |
| $q = 1$  | 0.3185     | 0.2268       | 0.1338       | 0.0940       | 0.0724       |
| $q = 0$  | 0.4997     | 0.2809       | 0.1484       | 0.1007       | 0.0762       |
| $q = -1$ | 0.7396     | 0.3428       | 0.1640       | 0.1076       | 0.0801       |

When the LDE coefficients of a $K = 2$ smoother in (8.2.3) are substituted into $H(\omega)$ in (8.1.5) and the integral in (8.4.1) in computed, the following closed-form expression for the WNG is obtained:

$$\text{WNG} = (1-p)\left[\frac{1}{(1+p)} + \frac{2d_\sigma}{(1+p)^2} + \frac{2d_\sigma^2}{(1+p)^3}\right] \text{ with} \tag{8.4.4a}$$

$$d_\sigma = p + pq - q. \tag{8.4.4b}$$

The optimal lag that minimizes the WNG is determined by evaluating its derivative (w.r.t $q$), setting it to zero, and solving for $q$, yielding the following closed-form expression for $q_{\text{opt}}$ as a function of $p$:

$$q_{\text{opt}} = \tfrac{1}{2}(1 + 3p)/(1 - p). \tag{8.4.5}$$

Using this lag places a zero at $z = -1$ for a magnitude null at $\omega = \pi$ in the frequency response. Optimal lags for the smoothers in Table 1 and the achieved WNG are shown in Table 2. Over this range of smoothing memories, using $q = 2l$ is a reasonable approximation, provided long system latencies are tolerable.

*Table 2 – Minimum white-noise gain (WNG) for the smoothers in Table 1 with an optimal lag.*

|               | $l = 2$      | $l = 4$      | $l = 8$      | $l = 12$     | $l = 16$     |
|---------------|--------------|--------------|--------------|--------------|--------------|
|               | $p = 0.6065$ | $p = 0.7788$ | $p = 0.8825$ | $p = 0.9200$ | $p = 0.9394$ |
| $q_{\text{opt}}$ | 3.58      | 7.54         | 15.52        | 23.51        | 31.51        |
| WNG           | 0.1225       | 0.0622       | 0.0312       | 0.0208       | 0.0156       |

## 8.5 Near-dc magnitude-and-phase linearity (flatness)

The smoothers considered here are designed to accommodate process models via an observer structure; and as such, they are not designed according to traditional 'passband' and 'stopband', specifications. For integrating systems, it is the asymptotic response at the $\omega \to 0$ limit that is important, or in the 'near-dc region' when non-infinite observer memory (in the time domain) and non-vanishing observer bandwidth (in the frequency domain) are considered.

The frequency response of the smoothers in Figure 6 is plotted in Figure 7. The magnitude attenuation and phase lag away from the near-dc region both increase with $p$ & $q$. A detail of the near-dc region is also shown in Figure 8. The effect of the $q$ parameter on the phase at very low frequencies is more apparent in these plots. As the filter memory $l$ decreases: the filter response is matched to the 'ideal' response over a wider range of frequencies, i.e. a higher bandwidth for lower bias; however, this is at the expense of higher WNG for higher variance (as shown in Table 1).

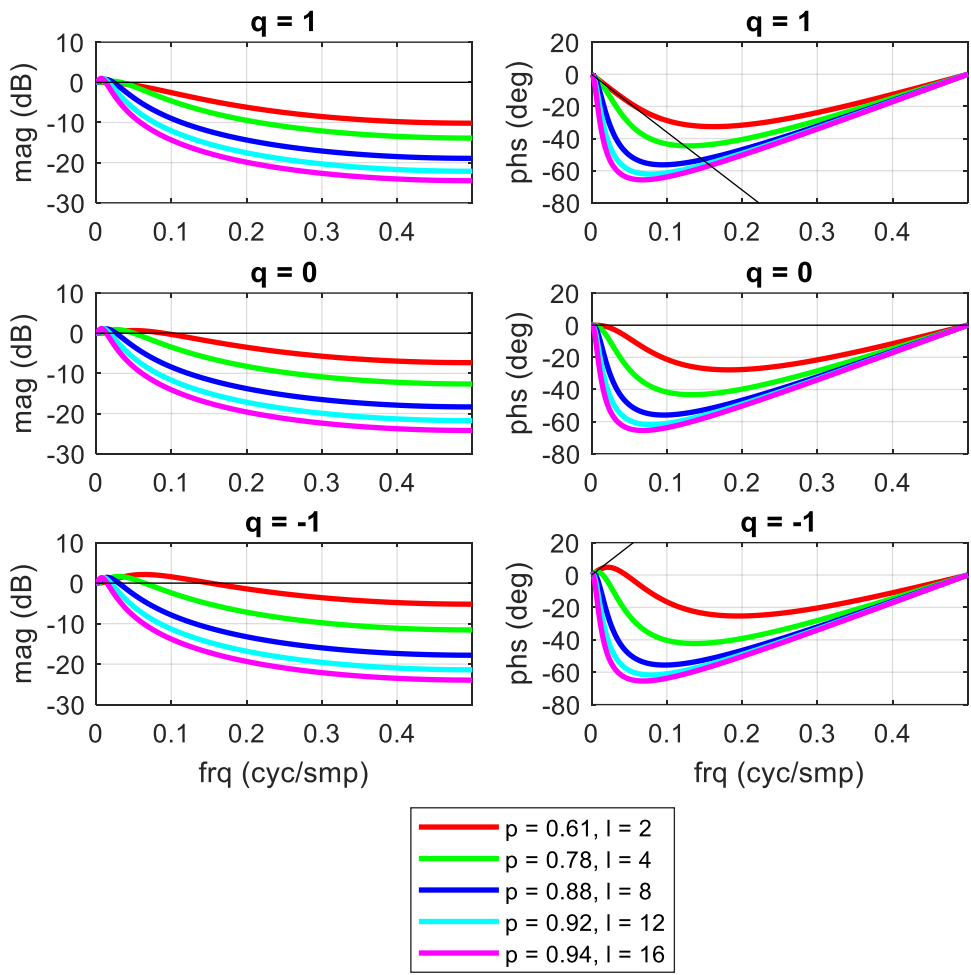

*Figure 7 – Frequency response of various second-order smoothers ($K = 2$). Magnitude response (left) and phase response (right) as a function of the normalized frequency ($f = \omega/2\pi$, cycles per sample). The observers are parameterized using delays ($q$) of 1, 0 and -1, samples (top to bottom) and five different pole locations ($p$). The poles are set using $p = e^{-1/l}$ where $l$ is the observer memory, in samples (see legend for values). Ideal low-frequency response for the smoothers is shown (black line). See Figure 8 for details of the near-dc response.*

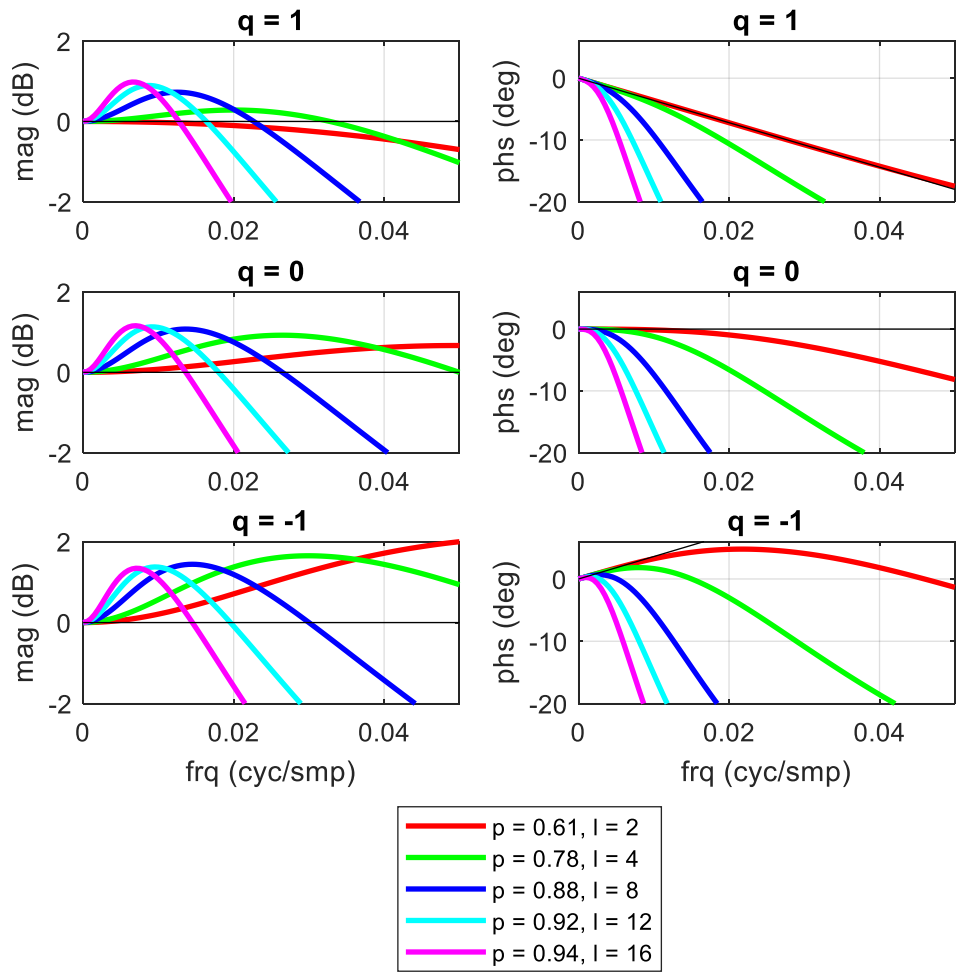

*Figure 8 – Near-dc frequency response of various second-order smoothers ($K = 2$). Magnitude response (left) and phase response (right) as a function of the normalized frequency ($f = \omega/2\pi$, cycles per sample). The observers are parameterized using delays ($q$) of 1, 0 and -1, samples (top to bottom) and five different pole locations ($p$). The poles are set using $p = e^{-1/l}$ where $l$ is the observer memory, in samples (see legend for values). Ideal low-frequency response for the smoothers is shown (black line). See Figure 7 for the full frequency range.*

When an estimate of the full (kinematic) state vector of a $K_t$th-order integrating process (with $K_t > 0$) is not required, an observer structure is unnecessary. The desired smoothers and differentiators may instead be designed directly in the frequency domain. A filter to estimate the $k_t$th element of the state vector (for $0 \leq k_t < K_t$), i.e. the $k_t$th derivative of the position w.r.t time, is simply obtained via a $K_\omega$th-order Taylor-series expansion (with $K_\omega > 0$) about $\omega = 0$. Matching each of the $k_\omega$th derivatives of the realized filter's frequency response $H(\omega)$, and the ideal differentiator's frequency response $H_{k_t}(\omega)$, for so-called 'flatness' in the frequency domain, ensures that the filter has the required number of so-called 'vanishing moments' in the time domain, for an unbiased estimate of the $k_t$th kinematic state (for $0 \leq k_\omega < K_\omega$ and $K_\omega = K_t = K$). The order of flatness and the number of vanishing moments is equal to $K$. Using a high-order state vector ($K_t \gg 0$) increases the filter bandwidth and the white-noise gain. Furthermore, using a high-order Taylor series ($K_\omega > K_t$) is unnecessary because an ideal differentiator does not have the properties required of a practical estimator. A filter with non-negligible bandwidth (in the $\omega$-domain) is required for non-infinite memory (in the $t$-domain) and a filter with a non-vanishing memory (in the $t$-domain) is required for adequate white-noise attenuation (in the $\omega$-domain). Thus, the frequency response of the filter away from the near-dc region should be shaped arbitrarily to meet other design requirements. Some frequency-domain design procedures for causal IIR filters in the time-domain are considered in [20] and [21]. Non-causal IIR and FIR filters for processing in the spatial domains (e.g. 2-D images) are considered in [22]. The causal IIR case is considered below.

The Laplace transform of an ideal $k_t$th-order differentiator (w.r.t time) is $s^{k_t}$ (for $0 \leq k_t$) thus its continuous-time frequency-response is found by substituting $s = i\Omega$, yielding

$$H_{k_t}(\Omega) = \mathcal{H}_{k_t}(s)\big|_{s=i\Omega} = s^{k_t}\big|_{s=i\Omega} = (i\Omega)^{k_t}. \tag{8.5.1}$$

The corresponding discrete-time frequency-response is then found by substituting $\Omega = \omega/T_s$, yielding

$$H_{k_t}(\omega) = H_{k_t}(\Omega)\big|_{\Omega=\omega/T_s} = (i\Omega)^{k_t}\big|_{\Omega=\omega/T_s} = \left(\frac{i\omega}{T_s}\right)^{k_t}. \tag{8.5.2}$$

The discrete-time transfer-function of a $q$-sample delay is $z^{-q}$. Its discrete-time frequency-response is found by substituting $z = e^{i\omega}$, yielding

$$H_q(\omega) = \mathcal{H}_q(z)\big|_{z=e^{i\omega}} = z^{-q}\big|_{z=e^{i\omega}} = e^{-iq\omega}. \tag{8.5.3}$$

The discrete-time frequency response of an ideal differentiator with a delay of $q$ samples is therefore

$$H_{q,k_t}(\omega) = H_q(\omega)H_{k_t}(\omega) = e^{-iq\omega}\left(\frac{i\omega}{T_s}\right)^{k_t}.$$

The $k_\omega$th derivative of this discrete-time frequency-response, evaluated at $\omega = 0$ is reached via $k_\omega$ applications of the product rule, yielding

$$\left\{\frac{d^{k_\omega}}{d\omega^{k_\omega}}H_{q,k_t}(\omega)\right\}\bigg|_{\omega=0} = \begin{cases} 0 & \text{for } k_\omega < k_t \\ i^{k_\omega}(-q)^{k_\omega-k_t}\left(\frac{1}{T_s}\right)^{k_t}\frac{k_\omega!}{(k_\omega-k_t)!} & \text{for } k_\omega \geq k_t \end{cases}. \tag{8.16}$$

The derivatives (w.r.t frequency and evaluated at dc) of the *realized* filter (for $0 \leq k_\omega < K_\omega$ with $K_\omega = K_t$) are matched to the corresponding derivatives of the *ideal* differentiator ($k_t > 0$) or smoother ($k_t = 0$). At frequencies away from dc, the response is shaped to meet other design objectives, e.g. the bias versus variance trade-off in target trackers, which distorts higher order derivatives (for $k_\omega > K_t$) away from their ideal values. Examples of other design objectives are: to attenuate high-frequency or coloured noise [20],[21]; to set the scale of image analysis [23], or for an isotropic magnitude response in steerable 2-D filters [22]. A different initialization procedure is required for these filters that are designed in the frequency domain using flatness constraints, i.e. without a process model. Start-up transients are suppressed by applying the final-value theorem in (8.3.1). The internal filter states are set equal to the steady-state values expected for a step input with a magnitude equal to the first input [24].

An analogue circuit or a digital circuit is used to physically instantiate a continuous-time or discrete-time LSS model in an engineered (i.e. synthetic) system, respectively. Coordinate transforms are usually applied to yield a model that it physically realizable, i.e. using integration operations instead of differentiation operations in analogue systems or using past samples instead of future samples in digital systems. They may also be used to reduce the number of components or operations required to build the system. These alternative 'canonical' forms reveal the poles and zeros of the corresponding transfer function, which allows the transient and steady-state response of the system to be determined and analyzed, for instance using its frequency response or impulse response. In the discrete-time case, the elements of the state-vector no longer have any obvious physical significance in this form and they simply represent the numerical contents of a delay register. At this point in the design cycle, it becomes obvious that a tracker (at steady-state), estimator, observer, or controller, incorporating an LTI model is simply a linear filter, and that the model is simply the means by which a filter with the desired properties is reached. A filter with the required impulse response and frequency response is designed from a process model and the process model is derived from empirical impulse-response and frequency-response data. If the properties of the required filter can be derived by other means, then process models (and pole placement) are not required [20],[21],[22].

# 9.    Closing remarks

Tracker design is an exercise is compromise – an improvement in one aspect of performance (e.g. increased bandwidth for reduced transient bias) necessarily degrades another (e.g. increased steady-state variance), for a perfectly modelled process. The frequency response is simple way of visualizing and quantifying this trade-off. However, it is rarely used as a design tool because it is usually assumed that the gains of a Kalman filter are optimal for a set of prior distributions and that nothing more therefore needs to be done. In this tutorial the utility of the frequency response is discussed – how to evaluate it, how to interpret it and how to shape it to meet performance specifications and overall behaviour. The Luenberger state observer (designed via pole placement) is recommended as a simple way of designing tracking filters when prior models for optimal Bayesian methods are either unknown or too difficult to work with (e.g. non-Gaussian) for online operation in real-time systems.

The design of fixed-gain state-observers by pole placement is a powerful tool because it allows the desired balance between the transient response and the steady-state response of a tracking filter to be set using a single smoothing parameter ($p$) between zero and one. A secondary lag parameter ($q$) is used to control the phase shift of the output (i.e. to propagate the states forward or backward in time by $q$ samples). Similar responses may also be achieved using a steady-state Kalman filter; indeed, the response is optimal for a given combination of process-noise and measurement-noise parameters; however, the assumed statistical distributions are rarely a reasonable approximation of operational reality; therefore, the designer of a tracking system is forced to empirically tune the tracker for the desired transient and steady-state response by manually adjusting the statistical parameters for an appropriate bandwidth so that the tracker requirements are met and the system operators are satisfied. Fixed-gain trackers designed by pole placement obviate the need for noise statistics and focus on system requirements/expectations directly. The performance may not be mathematically optimal, but the behaviour is always the same, regardless of the conditions.

# 11. APPENDIX A: Worked example

The primary purpose of this worked example is to illustrate the computation of the gain vector and the associated transformations required for response analysis and fast realization. Consider a state observer for a triple integrator (constant acceleration) target process. The output of the process is sampled at a frequency of $F_s = 25$ Hz (for $T_s = 0.04$ seconds). The observer is designed using $p = 0.8$ (for a memory of $l = 4.4814$ samples) and a lag of $q = 2$ (samples).

For this third-order process ($K = 3$), as specified in (3.1)

$$\boldsymbol{A}_{\text{prc}} = \begin{bmatrix} 0 & 1 & 0 \\ 0 & 0 & 1 \\ 0 & 0 & 0 \end{bmatrix} \text{ and } \boldsymbol{C}_{\text{prc}} = [1 \quad 0 \quad 0].$$

Then, following the procedure in (3.2a)-(3.2f), or simply using the end result in (3.2g) & (3.2h), yields

$$\boldsymbol{G}_{\text{prc}}^{\text{kin}} = \begin{bmatrix} 1 & T_s & \frac{T_s^2}{2} \\ 0 & 1 & T_s \\ 0 & 0 & 1 \end{bmatrix}$$

$$\begin{aligned}
&= \begin{bmatrix} 1.0000 & 0.0400 & 0.0008 \\ 0.0000 & 1.0000 & 0.0400 \\ 0.0000 & 0.0000 & 1.0000 \end{bmatrix} \text{ and} \\
\boldsymbol{C}_{\text{prc}}^{\text{kin}} &= \begin{bmatrix} 1 & 0 & 0 \end{bmatrix}
\end{aligned}$$

for use in (3.3). For our observer in (4.1), we have

$$\begin{aligned}
\boldsymbol{C}_{\text{prd}}^{\text{kin}} &= \boldsymbol{C}_{\text{prc}}^{\text{kin}} \boldsymbol{G}_{\text{prc}}^{\text{kin}} \\
&= \begin{bmatrix} 1 & 0 & 0 \end{bmatrix} \begin{bmatrix} 1.0000 & 0.0400 & 0.0008 \\ 0.0000 & 1.0000 & 0.0400 \\ 0.0000 & 0.0000 & 1.0000 \end{bmatrix} \\
&= \begin{bmatrix} 1.0000 & 0.0400 & 0.0008 \end{bmatrix}
\end{aligned}$$

as defined in (4.4), and for $q = 2$, we have, with the definitions is (4.2) and (4.3):

$$\begin{aligned}
\boldsymbol{C}_{\text{obs}}^{\text{kin}} &= \boldsymbol{C}_{\text{prc}}^{\text{kin}} \boldsymbol{G}_{\text{prc}}^{\text{kin}}(2) = \boldsymbol{C}_{\text{prc}}^{\text{kin}}\{\boldsymbol{G}_{\text{prc}}^{\text{kin}}\}^{-2} = \boldsymbol{C}_{\text{prc}}^{\text{kin}}\{\boldsymbol{G}_{\text{prc}}^{\text{kin}}\}^{-1}\{\boldsymbol{G}_{\text{prc}}^{\text{kin}}\}^{-1} \\
&= \begin{bmatrix} 1 & 0 & 0 \end{bmatrix} \begin{bmatrix} 1.0000 & -0.0400 & 0.0008 \\ 0.0000 & 1.0000 & -0.0400 \\ 0.0000 & 0.0000 & 1.0000 \end{bmatrix} \begin{bmatrix} 1.0000 & -0.0400 & 0.0008 \\ 0.0000 & 1.0000 & -0.0400 \\ 0.0000 & 0.0000 & 1.0000 \end{bmatrix} \\
&= \begin{bmatrix} 1.0000 & -0.0800 & 0.0032 \end{bmatrix}
\end{aligned}$$

which uses the fact that

$$\{\boldsymbol{G}_{\text{prc}}^{\text{kin}}\}^{-1} = \begin{bmatrix} 1 & T_s & \frac{T_s^2}{2} \\ 0 & 1 & T_s \\ 0 & 0 & 1 \end{bmatrix}^{-1} = \begin{bmatrix} 1 & -T_s & \frac{T_s^2}{2} \\ 0 & 1 & -T_s \\ 0 & 0 & 1 \end{bmatrix}$$

i.e. one-step-behind 'retrodiction'. For our process, that we have arbitrarily defined for our system/signal, using (6.9a) with $\rho_k = 1$ for all $k$, we know that

$\mathcal{A}_{\text{prc}}(z) = \prod_{k=0}^{K-1}(z - \rho_k) = (z - 1)^3 = z^3 - 3z^2 + 3z - 1$ thus
$\boldsymbol{a}_{\text{prc}} = \begin{bmatrix} 1 & -3 & 3 & -1 \end{bmatrix}$ and using (6.9b) we have

$$\boldsymbol{g}_{\text{prc}} = \begin{bmatrix} 1 \\ -3 \\ 3 \end{bmatrix} \text{ for }$$

$$\boldsymbol{G}_{\text{prc}}^{\text{pcf}} = \begin{bmatrix} 0 & 0 & 1 \\ 1 & 0 & -3 \\ 0 & 1 & 3 \end{bmatrix} \text{ and } \boldsymbol{C}_{\text{prd}}^{\text{pcf}} = \begin{bmatrix} 0 & 0 & 1 \end{bmatrix}$$

in PCF coordinates for the $\langle \boldsymbol{C}_{\text{prd}}^{\text{kin}}, \boldsymbol{G}_{\text{prc}}^{\text{kin}} \rangle$ pair (by definition).

For our observer, with poles that we have arbitrarily defined for the desired transient and steady-state response, using (6.8a) with $\lambda_k = p = 0.8$ for all $k$, we know that

$\mathcal{A}_{\text{obs}}(z) = \prod_{k=0}^{K-1}(z - \lambda_k) = (z - 0.8)^3 = z^3 - 2.400z^2 + 1.920z - 0.512$ thus
$\boldsymbol{a}_{\text{obs}} = \begin{bmatrix} 1.000 & -2.400 & 1.920 & -0.512 \end{bmatrix}$ and using (6.8b) we have

$$\boldsymbol{g}_{\text{obs}} = \begin{bmatrix} 0.512 \\ -1.920 \\ 2.400 \end{bmatrix} \text{ for}$$

$$\boldsymbol{G}_{\text{obs}}^{\text{pcf}} = \begin{bmatrix} 0.0000 & 0.0000 & 0.512 \\ 1.0000 & 0.0000 & -1.920 \\ 0.0000 & 1.0000 & 2.400 \end{bmatrix}$$

in PCF coordinates for the $\langle \boldsymbol{C}_{\text{prd}}^{\text{kin}}, \boldsymbol{G}_{\text{prc}}^{\text{kin}} \rangle$ pair ($\boldsymbol{C}_{\text{obs}}^{\text{pcf}}$ is not yet known). Using (6.7) with (6.1)-(6.6) the gain vector in PCF coordinates is found using

$$\begin{aligned} \mathcal{K}^{\text{pcf}} &= \boldsymbol{g}_{\text{prc}} - \boldsymbol{g}_{\text{obs}} \\ &= \begin{bmatrix} 1 \\ -3 \\ 3 \end{bmatrix} - \begin{bmatrix} 0.512 \\ -1.920 \\ 2.400 \end{bmatrix} \\ &= \begin{bmatrix} 0.488 \\ -1.080 \\ 0.600 \end{bmatrix}. \end{aligned}$$

## 11.1 KIN

We now need to find the transform that converts this gain vector back into kinematic coordinates. The observability matrices in both coordinate systems are

$$\mathcal{O}_{\text{prc}}^{\text{kin}} = \begin{bmatrix} \boldsymbol{C}_{\text{prd}}^{\text{kin}} \{\boldsymbol{G}_{\text{prc}}^{\text{kin}}\}^0 \\ \boldsymbol{C}_{\text{prd}}^{\text{kin}} \{\boldsymbol{G}_{\text{prc}}^{\text{kin}}\}^1 \\ \boldsymbol{C}_{\text{prd}}^{\text{kin}} \{\boldsymbol{G}_{\text{prc}}^{\text{kin}}\}^2 \end{bmatrix}$$

$$= \begin{bmatrix} [1.0000 \;\; 0.0400 \;\; 0.0008] \begin{bmatrix} 1 & 0 & 0 \\ 0 & 1 & 0 \\ 0 & 0 & 1 \end{bmatrix} \\ [1.0000 \;\; 0.0400 \;\; 0.0008] \begin{bmatrix} 1.0000 & 0.0400 & 0.0008 \\ 0.0000 & 1.0000 & 0.0400 \\ 0.0000 & 0.0000 & 1.0000 \end{bmatrix} \\ [1.0000 \;\; 0.0400 \;\; 0.0008] \begin{bmatrix} 1.0000 & 0.0400 & 0.0008 \\ 0.0000 & 1.0000 & 0.0400 \\ 0.0000 & 0.0000 & 1.0000 \end{bmatrix} \begin{bmatrix} 1.0000 & 0.0400 & 0.0008 \\ 0.0000 & 1.0000 & 0.0400 \\ 0.0000 & 0.0000 & 1.0000 \end{bmatrix} \end{bmatrix}$$

$$= \begin{bmatrix} 1.0000 & 0.0400 & 0.0008 \\ 1.0000 & 0.0800 & 0.0032 \\ 1.0000 & 0.1200 & 0.0072 \end{bmatrix}$$

and

$$\mathcal{O}_{\text{prc}}^{\text{pcf}} = \begin{bmatrix} \boldsymbol{C}_{\text{prd}}^{\text{pcf}} \{\boldsymbol{G}_{\text{prc}}^{\text{pcf}}\}^0 \\ \boldsymbol{C}_{\text{prd}}^{\text{pcf}} \{\boldsymbol{G}_{\text{prc}}^{\text{pcf}}\}^1 \\ \boldsymbol{C}_{\text{prd}}^{\text{pcf}} \{\boldsymbol{G}_{\text{prc}}^{\text{pcf}}\}^2 \end{bmatrix} = \begin{bmatrix} [0 \;\; 0 \;\; 1] \begin{bmatrix} 1 & 0 & 0 \\ 0 & 1 & 0 \\ 0 & 0 & 1 \end{bmatrix} \\ [0 \;\; 0 \;\; 1] \begin{bmatrix} 0 & 0 & 1 \\ 1 & 0 & -3 \\ 0 & 1 & 3 \end{bmatrix} \\ [0 \;\; 0 \;\; 1] \begin{bmatrix} 0 & 0 & 1 \\ 1 & 0 & -3 \\ 0 & 1 & 3 \end{bmatrix} \begin{bmatrix} 0 & 0 & 1 \\ 1 & 0 & -3 \\ 0 & 1 & 3 \end{bmatrix} \end{bmatrix} = \begin{bmatrix} 0 & 0 & 1 \\ 0 & 1 & 3 \\ 1 & 3 & 6 \end{bmatrix}.$$

The required transform is then found using (6.11)

$$\mathbb{T}_{\text{prc}}^{\text{kin}\leftarrow\text{pcf}} = \{\mathcal{O}_{\text{prc}}^{\text{kin}}\}^{-1} \mathcal{O}_{\text{prc}}^{\text{pcf}}$$

$$
= \begin{bmatrix} 1.0000 & 0.0400 & 0.0008 \\ 1.0000 & 0.0800 & 0.0032 \\ 1.0000 & 0.1200 & 0.0072 \end{bmatrix}^{-1} \begin{bmatrix} 0 & 0 & 1 \\ 0 & 1 & 3 \\ 1 & 3 & 6 \end{bmatrix}
$$

$$
= 10^3 \times \begin{bmatrix} 0.0030 & -0.0030 & 0.0010 \\ -0.0625 & 0.1000 & -0.0375 \\ 0.6250 & -1.2500 & 0.6250 \end{bmatrix} \begin{bmatrix} 0 & 0 & 1 \\ 0 & 1 & 3 \\ 1 & 3 & 6 \end{bmatrix}
$$

$$
= \begin{bmatrix} 1.0 & 0.0 & 0.0 \\ -37.5 & -12.5 & 12.5 \\ 625.0 & 625.0 & 625.0 \end{bmatrix}
$$

and

$$
\mathbb{T}_{\text{prc}}^{\text{pcf}\leftarrow\text{kin}} = \left\{ \mathbb{T}_{\text{prc}}^{\text{kin}\leftarrow\text{pcf}} \right\}^{-1}
$$

$$
= \begin{bmatrix} 1.0 & 0.0 & 0.0 \\ -37.5 & -12.5 & 12.5 \\ 625.0 & 625.0 & 625.0 \end{bmatrix}^{-1} = \begin{bmatrix} 1.0 & 0.00 & 0.0000 \\ -2.0 & -0.04 & 0.0008 \\ 1.0 & 0.04 & 0.0008 \end{bmatrix}.
$$

The gain vector in kinematic coordinates is then found using (6.10), i.e.

$$
\boldsymbol{\mathcal{K}}^{\text{kin}} = \mathbb{T}_{\text{prc}}^{\text{kin}\leftarrow\text{pcf}} \boldsymbol{\mathcal{K}}^{\text{pcf}}
$$

$$
= \begin{bmatrix} 1.0 & 0.0 & 0.0 \\ -37.5 & -12.5 & 12.5 \\ 625.0 & 625.0 & 625.0 \end{bmatrix} \begin{bmatrix} 0.488 \\ -1.080 \\ 0.600 \end{bmatrix} = \begin{bmatrix} 0.4880 \\ 2.7000 \\ 5.0000 \end{bmatrix}.
$$

Substitution of $\boldsymbol{\mathcal{K}}^{\text{kin}}$ into (6.1c) yields

$$
\boldsymbol{G}_{\text{obs}}^{\text{kin}} = \boldsymbol{G}_{\text{prc}}^{\text{kin}} - \boldsymbol{\mathcal{K}}^{\text{kin}} \boldsymbol{C}_{\text{prd}}^{\text{kin}}
$$

$$
= \begin{bmatrix} 1.0000 & 0.0400 & 0.0008 \\ 0.0000 & 1.0000 & 0.0400 \\ 0.0000 & 0.0000 & 1.0000 \end{bmatrix} - \begin{bmatrix} 0.4880 \\ 2.7000 \\ 5.0000 \end{bmatrix} \begin{bmatrix} 1.0000 & 0.0400 & 0.0008 \end{bmatrix}
$$

$$
= \begin{bmatrix} 0.5120 & 0.0205 & 0.0004 \\ -2.7000 & 0.8920 & 0.0378 \\ -5.0000 & -0.2000 & 0.9960 \end{bmatrix} \text{ and}
$$

$$
\boldsymbol{H}_{\text{obs}}^{\text{kin}} = \boldsymbol{\mathcal{K}}^{\text{kin}}
$$

$$
= \begin{bmatrix} 0.4880 \\ 2.7000 \\ 5.0000 \end{bmatrix}.
$$

We now have everything we need to realize our observer, using the block diagram in Figure 5, with $\langle \boldsymbol{C}_{\text{obs}}^{\text{kin}}, \boldsymbol{G}_{\text{obs}}^{\text{kin}}, \boldsymbol{H}_{\text{obs}}^{\text{kin}} \rangle$, as defined in Section 7.1.

## 11.2 OCF

The transform for an OCF is obtained using the observability matrices as follows:

$$
\mathcal{O}_{\text{obs}}^{\text{kin}} = \begin{bmatrix} \boldsymbol{C}_{\text{obs}}^{\text{kin}} \{\boldsymbol{G}_{\text{obs}}^{\text{kin}}\}^0 \\ \boldsymbol{C}_{\text{obs}}^{\text{kin}} \{\boldsymbol{G}_{\text{obs}}^{\text{kin}}\}^1 \\ \boldsymbol{C}_{\text{obs}}^{\text{kin}} \{\boldsymbol{G}_{\text{obs}}^{\text{kin}}\}^2 \end{bmatrix}
$$

$$
\begin{aligned}
&= \begin{bmatrix} [1.0000 & -0.0800 & 0.0032] \begin{bmatrix} 1 & 0 & 0 \\ 0 & 1 & 0 \\ 0 & 0 & 1 \end{bmatrix} \\ [1.0000 & -0.0800 & 0.0032] \begin{bmatrix} 0.5120 & 0.0205 & 0.0004 \\ -2.7000 & 0.8920 & 0.0378 \\ -5.0000 & -0.2000 & 0.9960 \end{bmatrix} \\ [1.0000 & -0.0800 & 0.0032] \begin{bmatrix} 0.5120 & 0.0205 & 0.0004 \\ -2.7000 & 0.8920 & 0.0378 \\ -5.0000 & -0.2000 & 0.9960 \end{bmatrix} \begin{bmatrix} 0.5120 & 0.0205 & 0.0004 \\ -2.7000 & 0.8920 & 0.0378 \\ -5.0000 & -0.2000 & 0.9960 \end{bmatrix} \end{bmatrix} \\
&= \begin{bmatrix} 1.0000 & -0.0800 & 0.0032 \\ 0.7120 & -0.0515 & 0.0006 \\ 0.5008 & -0.0315 & -0.0011 \end{bmatrix} \text{ and}
\end{aligned}
$$

$$
\mathcal{O}_{\text{obs}}^{\text{ocf}} = \begin{bmatrix} \boldsymbol{C}_{\text{obs}}^{\text{ocf}} \{\boldsymbol{G}_{\text{obs}}^{\text{ocf}}\}^0 \\ \boldsymbol{C}_{\text{obs}}^{\text{ocf}} \{\boldsymbol{G}_{\text{obs}}^{\text{ocf}}\}^1 \\ \boldsymbol{C}_{\text{obs}}^{\text{ocf}} \{\boldsymbol{G}_{\text{obs}}^{\text{ocf}}\}^2 \end{bmatrix} = \begin{bmatrix} [0 & 0 & 1] \begin{bmatrix} 1 & 0 & 0 \\ 0 & 1 & 0 \\ 0 & 0 & 1 \end{bmatrix} \\ [0 & 0 & 1] \begin{bmatrix} 0 & 0 & 0.512 \\ 1 & 0 & -1.920 \\ 0 & 1 & 2.400 \end{bmatrix} \\ [0 & 0 & 1] \begin{bmatrix} 0 & 0 & 0.512 \\ 1 & 0 & -1.920 \\ 0 & 1 & 2.400 \end{bmatrix} \begin{bmatrix} 0 & 0 & 0.512 \\ 1 & 0 & -1.920 \\ 0 & 1 & 2.400 \end{bmatrix} \end{bmatrix} = \begin{bmatrix} 0 & 0 & 1 \\ 0 & 1 & 2.40 \\ 1 & 2.40 & 3.84 \end{bmatrix}.
$$

The required transforms are then found using (7.3.1d) & (7.3.1e)

$$
\begin{aligned}
\mathbb{T}_{\text{obs}}^{\text{kin}\leftarrow\text{ocf}} &= \{\mathcal{O}_{\text{obs}}^{\text{kin}}\}^{-1} \mathcal{O}_{\text{obs}}^{\text{ocf}} \\
&= \begin{bmatrix} 1.0000 & -0.0800 & 0.0032 \\ 0.7120 & -0.0515 & 0.0006 \\ 0.5008 & -0.0315 & -0.0011 \end{bmatrix}^{-1} \begin{bmatrix} 0 & 0 & 1 \\ 0 & 1 & 2.40 \\ 1 & 2.40 & 3.84 \end{bmatrix} \\
&= 10^6 \times \begin{bmatrix} 0.0181 & -0.0459 & 0.0291 \\ 0.2592 & -0.6575 & 0.4172 \\ 0.8256 & -2.0938 & 1.3281 \end{bmatrix} \begin{bmatrix} 0 & 0 & 1 \\ 0 & 1 & 2.40 \\ 1 & 2.40 & 3.84 \end{bmatrix} \\
&= 10^6 \times \begin{bmatrix} 0.0291 & 0.0240 & 0.0198 \\ 0.4172 & 0.3438 & 0.2832 \\ 1.3281 & 1.0938 & 0.9006 \end{bmatrix} \text{ and}
\end{aligned}
$$

$$
\begin{aligned}
\mathbb{T}_{\text{obs}}^{\text{ocf}\leftarrow\text{kin}} &= \{\mathbb{T}_{\text{obs}}^{\text{kin}\leftarrow\text{ocf}}\}^{-1} \\
&= \left\{ 10^6 \times \begin{bmatrix} 0.0291 & 0.0240 & 0.0198 \\ 0.4172 & 0.3438 & 0.2832 \\ 1.3281 & 1.0938 & 0.9006 \end{bmatrix} \right\}^{-1} = \begin{bmatrix} 0.7120 & -0.0614 & 0.0037 \\ -1.6880 & 0.1405 & -0.0071 \\ 1.0000 & -0.0800 & 0.0032 \end{bmatrix}.
\end{aligned}
$$

Finally, with the assistance of (7.3.2c), the $\boldsymbol{b}_{\text{obs}}$ coefficients are extracted from $\boldsymbol{H}_{\text{obs}}^{\text{ocf}}$ as follows:

$$
\begin{aligned}
\boldsymbol{H}_{\text{obs}}^{\text{ocf}} &= \mathbb{T}_{\text{obs}}^{\text{ocf}\leftarrow\text{kin}} \boldsymbol{H}_{\text{obs}}^{\text{kin}} \\
&= \begin{bmatrix} 0.7120 & -0.0614 & 0.0037 \\ -1.6880 & 0.1405 & -0.0071 \\ 1.0000 & -0.0800 & 0.0032 \end{bmatrix} \begin{bmatrix} 0.4880 \\ 2.7000 \\ 5.0000 \end{bmatrix} \\
&= \begin{bmatrix} 0.2000 \\ -0.4800 \\ 0.2880 \end{bmatrix}. \text{ Thus, as indicated in (8.1.1):}
\end{aligned}
$$
$\boldsymbol{b}_{\text{obs}} = [0.288 \quad -0.480 \quad 0.200 \quad 0.000]$.

We now have everything we need to realize our observer, using the block diagram in Figure 5; with: $\mathbb{T}_{\text{obs}}^{\text{ocf}\leftarrow\text{kin}}$ for filter initialization, $\langle \boldsymbol{C}_{\text{obs}}^{\text{ocf}}, \boldsymbol{G}_{\text{obs}}^{\text{ocf}}, \boldsymbol{H}_{\text{obs}}^{\text{ocf}} \rangle$ for state propagation and output generation, along with $\mathbb{T}_{\text{obs}}^{\text{kin}\leftarrow\text{ocf}}$ for kinematic state extraction, as defined in Section 7.3.

Alternatively, the $\boldsymbol{b}_\text{obs}$ & $\boldsymbol{a}_\text{obs}$ coefficients may be passed to a standard filtering function for batch processing, if the full kinematic state is not required at each time step. However, to avoid erratic startup transients, care must be taken to appropriately initialize the internal state of the filter [24]. Furthermore, the final internal state of the filter must be transformed to produce the kinematic state of the estimator at the end of the batch. Therefore, it is important to understand how the registers of the standard filtering function are arranged.

# 12. APPENDIX B: A gentle introduction to the magic of linear time-invariant signals and systems

In this tutorial, it is assumed that the reader has a reasonable working knowledge of complex numbers and linear algebra, as acquired through advanced secondary or first-year tertiary studies. Unfortunately, it is also assumed that the reader has been exposed to some signals-and-systems theory (e.g. the Laplace transform) during undergraduate studies in engineering and that it made no sense at the time. It is also unfortunate that in this modern age, where the 'intelligent' machines we build are based on digital computers, that (continuous-time) $s$-plane analysis is considered an essential core yet (discrete-time) $z$-plane analysis (reached via the $\mathcal{Z}$-transform) is an optional extra. An attempt is made to remedy this situation in this appendix. The focus here is on concepts and principles. Standard signals-and-systems textbooks should be consulted for more detailed/rigorous proofs and derivations [8],[9].

## 12.1 From Fourier to Laplace and beyond

The one-dimensional Fourier domain is a vertical 'slice', along the imaginary $\Omega$-axis, through the two-dimensional Laplace domain (i.e. the complex $s$-plane).

The Fourier domain is used to represent the *steady-state* response of an LTI system to a sinusoidal input of unity magnitude and *infinite* duration. The frequency response $H(\Omega)$, of a system is a complex function of the *imaginary* frequency variable $i\Omega$ (for $-\infty < \Omega < \infty$, where $\Omega$ is real and in units of radians per second), thus the steady-state output of that system $y(t)$, to the complex input $x(t) = e^{i\Omega t}$, is simply expressed as a magnitude scaling and a phase shift, using $y(t) = H(\Omega)x(t)$.

The Laplace domain is used to represent the *transient* response of an LTI system to a more general oscillatory input of non-constant magnitude (exponentially growing or decaying) and *finite* duration (starting from zero time, i.e. at $t = 0$ seconds). The continuous-time transfer-function $\mathcal{H}(s)$, of a system is also a complex function of the now *complex* variable $s = \sigma + i\Omega$ (for $-\infty < \sigma < \infty$, where $\sigma$ is real and in units of reciprocal seconds). The output of the system, from a zero initial-state, is now found via an inverse Laplace transform using $y(t) = \mathcal{L}^{-1}\{\mathcal{H}(s)\mathcal{X}(s)\}$. This result follows from the definition $\mathcal{H}(s) = \mathcal{Y}(s)/\mathcal{X}(s)$ where $\mathcal{X}(s) = \mathcal{L}\{x(t)\}$ and $\mathcal{Y}(s) = \mathcal{L}\{y(t)\}$, i.e. the Laplace transform of the input and output, respectively.

The continuous-time transfer-function $\mathcal{H}(s)$, for a $K$th-order proper system, that links the system input to the system output, is expressed as a ratio of polynomials in $s$

$$\mathcal{H}(s) = \mathcal{B}(s)/\mathcal{A}(s) = \sum_{k=0}^{K} b[k]s^{K-k} / \sum_{k=0}^{K} a[k]s^{K-k} . \tag{12.1.1}$$

As $s^k$ is a $k$th-order derivative operation (w.r.t. time), the continuous-time transfer-function $\mathcal{H}(s)$, is a Laplace-domain representation of a linear differential equation. After dividing the numerator and denominator of (12.1.1) by $s^K$, $\mathcal{H}(s)$ is instead expressed using integrals, i.e.

$$\mathcal{H}(s) = \sum_{k=0}^{K} b[k]s^{-k} / \sum_{k=0}^{K} a[k]s^{-k} \tag{12.1.2}$$

where $1/s$ is the Laplace transform of an integral operator.

The $K$ (complex) poles of the system $\lambda_k$, i.e. the $K$ roots of the $K$th-degree denominator polynomial $\mathcal{A}(s)$, where $|\mathcal{H}(s)| \to \infty$, represent its natural resonant modes. For an impulse input, the output is a linear combination of these $K$ modes. The $K$ (complex) zeros of the system, i.e. the $K$ roots of the $K$th-degree numerator polynomial $\mathcal{B}(s)$, where $|\mathcal{H}(s)| \to 0$, are determined from these complex linear coefficients, which scale and shift each mode, in magnitude and phase, respectively. The coefficients $c[k]$, are found from a partial fraction expansion of 12.1.1, which changes the form of $\mathcal{H}(s)$, assuming it has no repeated poles, to

$$\mathcal{H}(s) = \sum_{k=0}^{K-1} c[k]/(s + \lambda_k). \tag{12.1.3}$$

The (horizontal) real axis of the complex $s$-plane therefore provides an additional degree of freedom that may be used to describe the form of a signal and the response of a system. The real part of the $s$ variable describes the rate of exponential decay or growth (i.e. the 'envelope') whereas the imaginary part describes the frequency of oscillation (i.e. the 'carrier') in the time domain. A simple example: For the continuous-time impulse-response $h(t) = e^{(\sigma + i\Omega)t}$ (for $t \geq 0$; with $h(t) = 0$ for $t < 0$) we have its continuous-time transfer-function $\mathcal{H}(s) = \mathcal{L}\{h(t)\} = 1/(s - \lambda)$, where $\lambda$ is the system pole at $s = \sigma + i\Omega$. When the impulse response is rewritten as $h(t) = e^{(\sigma + i\Omega)t} = e^{\sigma t}e^{i\Omega t}$ the respective roles of the real and imaginary parts (i.e. the envelope and carrier) are obvious. The left-hand side of the complex $s$-plane (where $\sigma < 0$) represents exponential decay (or stability) and the right-hand side of the complex $s$-plane (where $\sigma > 0$) represents exponential growth (or instability). Along the imaginary axis (where $\sigma = 0$) there is neither growth nor decay (so-called 'marginal' stability). As we move along the imaginary axis away from the origin, the frequency of oscillation increases.

The impulse response is an invaluable representation of continuous-time system. However, it requires us to accept that it is possible to synthesize an input of infinitesimal width and infinite amplitude (for unity area) i.e. the Dirac delta function. Thus, this system representation is a notional limiting case. Perhaps this situation is not too far removed from a frequency-response representation, where we are required to synthesize perfect sinusoidal inputs and to measure the system's output, once it has reached steady-state, after an infinite time has elapsed. The impulse response of a discrete-time system is a much less abstract concept as it is perfectly reasonable to synthesize a unit-impulse input, using a single value of one in a long sequence of zeros.

Indeed, all *s*-domain concepts have an analogue in the *z*-domain. However, they are arguably easier to understand in the latter domain because they rely less on the calculus of infinitesimals and limits and more on numeric values of finite precision at discrete moments in time. Moreover, discrete-time transfer-functions may be realized *exactly* in digital computers using cascaded delay registers, each with a transfer function of $1/z$ (along with arithmetic multiplies). Whereas, continuous-time transfer-functions must be *approximated* in analogue circuits, using cascaded integrating elements, each with an *ideal* transfer function of $1/s$ (along with power amplifiers). Furthermore, the behaviour of analogue circuits changes with time, e.g. as the device warms up after it is switched on, or as the components degrade with age. The mathematics of discrete-time systems is also much more accessible because it is very easy to determine the output of a system at a given point in time, from a given initial state, for a given input, exactly. Only a simple `for` loop is required, not a differential equation solver.

## 12.2 From the *s*-plane to the *z*-plane (or from solder to silicon)

When a continuous-time signal is sampled uniformly (at a rate of $F_s = 1/T_s$ samples per second or Hz) to form a discrete-time signal, there is clearly a loss of information due to the non-zero sampling period ($T_s$ in seconds), as the behaviour of the signal between the sampling times is unknown. Signal components ($e^{i\Omega t}$) with frequencies that are greater than half the sampling rate ($|\Omega| > \Omega_s/2$) are aliased and the ability to resolve transients ($e^{\sigma t}$) with rates of decay that are less than the sampling frequency ($\sigma \ll -F_s$) is diminished.

These limitations reveal the way in which the complex *s*-plane is warped to yield the complex *z*-plane using $z = e^{sT_s} = e^{(\sigma+i\Omega)T_s}$. In this new domain, Re($z$) and Im($z$) correspond to the (dimensionless) horizonal and vertical axes, respectively; however, the physical significance of a discrete-time transfer-function $\mathcal{H}(z)$, in the *z*-plane is best interpreted using a polar coordinate system. In going from the *s*-plane to the *z*-plane: A 'hard' limit is placed on the vertical $i\Omega$ axis in the *s* plane and that Cartesian coordinate, originally of infinite extent, is truncated and 'bent' until it wraps around to become the angular coordinate (i.e. the angular frequency $\omega$, in units of radians per sample) in the *z* plane; Similarly, a 'soft' limit is placed on the horizontal $\sigma$ axis in the *s* plane and that Cartesian coordinate, originally of infinite extent, is represented using the radial coordinate in the *z* plane, with very negative values of $\sigma$ in the *s* plane, corresponding to rapid exponential decay, 'squeezed' into the region around the origin of the *z* plane where $|z| \to 0$. The so-called 'unit circle' in the *z* plane where $|z| = 1$ corresponds to a pure oscillation, with no exponential growth or decay, thus the discrete-time frequency-response of a system is determined by evaluating the discrete-time transfer-function around this unit circle, using $H(\omega) = \mathcal{H}(z)|_{z=e^{i\omega}}$. Similarly, the region outside the unit circle where $|z| > 1$ corresponds to exponential growth.

The output of the system, from a zero initial-state, is now found via an inverse $\mathcal{Z}$ transform using $y[n] = \mathcal{Z}^{-1}\{\mathcal{H}(z)\mathcal{X}(z)\}$. This result follows from the definition $\mathcal{H}(z) = \mathcal{Y}(z)/\mathcal{X}(z)$ where $\mathcal{X}(z) = \mathcal{Z}\{x[n]\}$ and $\mathcal{Y}(z) = \mathcal{Z}\{y[n]\}$, i.e. the $\mathcal{Z}$ transform of the input and output, respectively. As in continuous-time, a *K*th-order (proper) discrete-time transfer-function $\mathcal{H}(z)$, is expressed as

$$\mathcal{H}(z) = \mathcal{B}(z)/\mathcal{A}(z) = \sum_{k=0}^{K} b[k]z^{K-k}/\sum_{k=0}^{K} a[k]z^{K-k} \qquad (12.2.1a)$$

or

$$\mathcal{H}(z) = \sum_{k=0}^{K} b[k]z^{-k}/\sum_{k=0}^{K} a[k]z^{-k} \qquad (12.2.1b)$$

where $1/z$ is the $\mathcal{Z}$ transform of a one-sample delay (i.e. the so-called 'unit' delay).

The relationship between the complex $s$-plane for a continuous-time first-order system (see Figure 9), the complex $z$-plane for a discrete-time first-order system (see Figure 10), their corresponding continuous-time and discrete-time impulse-responses (see Figure 11), and their corresponding discrete-time frequency-responses (see Figure 12) are shown below.

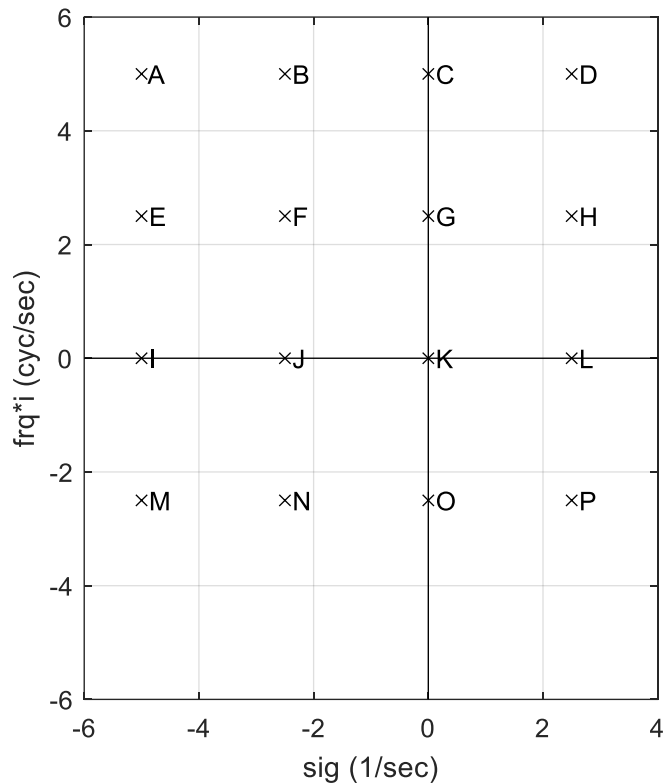

*Figure 9 – Poles of a continuous-time first-order system with impulse response $h(t) = e^{(\sigma + i\Omega)t}$ in the complex s-plane: rate of decay/growth (σ, in reciprocal seconds) on the horizontal axis, normalized imaginary frequency (iΩ/2π, in cycles per second) on the vertical axis. The system is parameterized using $\sigma = -1/\tau$ and $\Omega = 2\pi/\tau$, where τ is the coherence duration (in seconds), for $\tau \in \{0.2, 0.4, \infty, -0.4\}$ from left to right and top to bottom.*

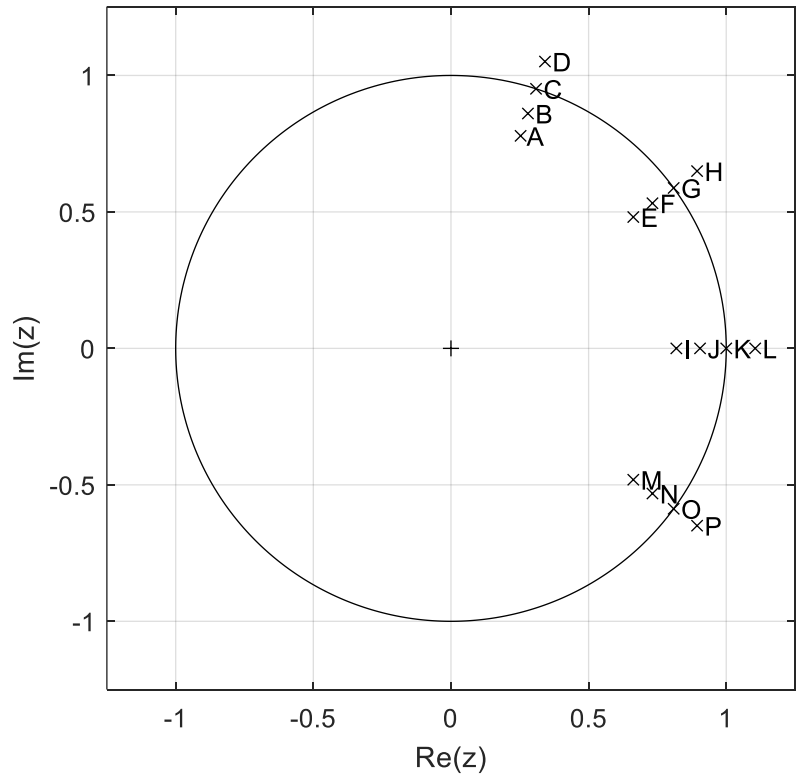

*Figure 10 - Poles of a discrete-time first-order system with impulse response $h[n] = e^{T_s(\sigma+i\Omega)n}$ in the complex z-plane, for $T_s = 1/25$ seconds. See Figure 9 for system parameters and the poles of the corresponding continuous-time system.*

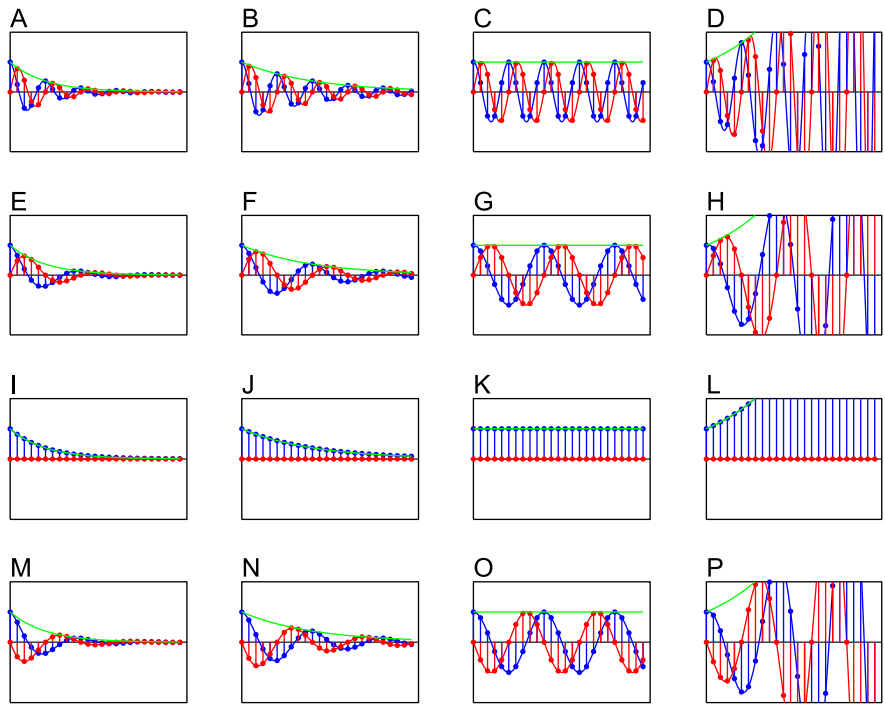

*Figure 11 - Impulse responses of the continuous-time and discrete-time first-order systems considered in Figure 9 and Figure 10 (respectively) over a time interval of 1 second and an amplitude range of $\pm 2$. See Figure 9 for system parameters. Real part of response in blue, imaginary part in red, magnitude in green. Note that the rows have the same carrier frequency whereas the columns have the same envelope. Note also that the imaginary parts cancel in second-order oscillatory systems with complex poles in conjugate pairs; however, for these first-order systems the response is complex.*

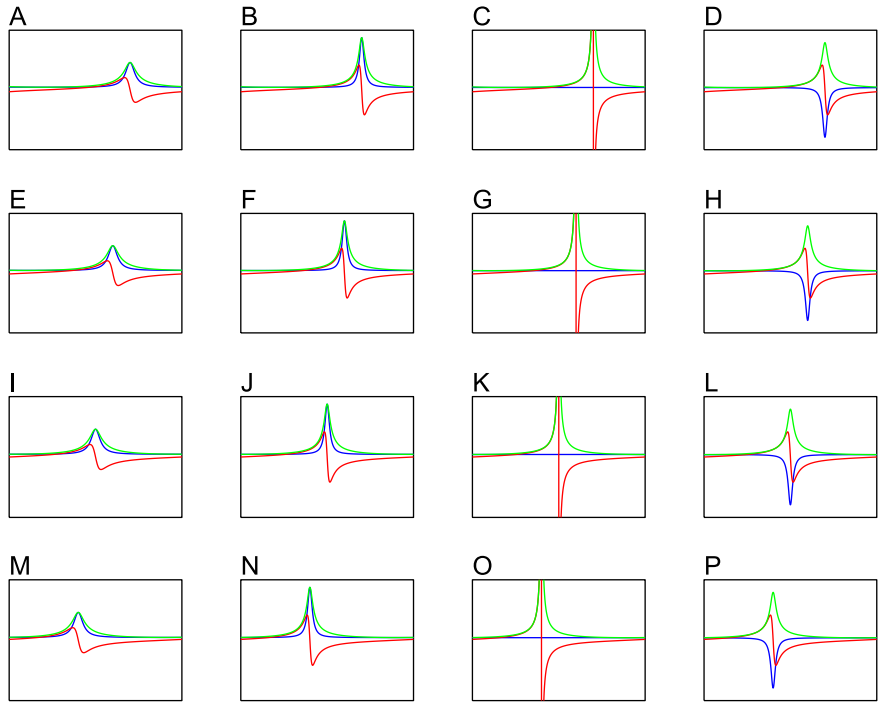

*Figure 12 - Frequency responses of the discrete-time first-order systems considered in Figure 10 over an angular frequency interval of $\omega \in \{-\pi, \pi\}$ and a response range of $\pm 12$. Real part of response in blue, imaginary part in red, magnitude in green. Note that the height of the gain (i.e. magnitude) peak increases as the pole approaches the unit circle. The peaks become singularities for poles on the unit circle.*

The impulse response for the $k = 0$ element (i.e. position, for a smoothing filter), output by the state observer derived in Section 11, is shown in Figure 13. A standard representation of the corresponding frequency response is shown in Figure 14. The phase and magnitude are computed from the complex response shown in Figure 15, which is determined by evaluating the discrete-time transfer-function around the unit circle, as shown in Figure 16. As the three poles of the observer are moved from $z = 0.8$ to $z = 0.0$, i.e. the origin of the $z$ plane, the discrete-time transfer-function simply becomes a delay of two samples with a frequency response of unity magnitude, i.e. $\boldsymbol{b} = [0 \quad 0 \quad 1 \quad 0]$ and $\boldsymbol{a} = [1 \quad 0 \quad 0 \quad 0]$, for one zero at $z = 0$ and three poles at $z = 0$, thus $\mathcal{H}(z) = z^{-2}$ and $H(\omega) = e^{-2i\omega}$. The corresponding plots for this pure-delay 'all-pass' observer, i.e. with no white-noise attenuation, are shown in Figure 17, Figure 18, Figure 19 and Figure 20.

The pure delay case above illustrates a tenet of signals-and-systems theory, a consequence of the (continuous-time) Fourier transform and its inverse, which should be acknowledged before proceeding:

An 'impulse' in the frequency domain represents a sinusoid in the time domain, or a complex-modulation operation in the time domain represents a 'shift' in the frequency domain. For example, a unit impulse at the centre of the $k$th bin of an $M$-sample DFT spectrum (for $-K \leq k \leq K$ and $M = 2K + 1$) corresponds to a signal $f_k[n] = e^{i\omega n}$ in the time domain with an angular frequency of $\omega = 2\pi k/M$ (radians per sample) where $t = nT_s$ and $n$ is the sample index. This concept is understood by anybody who has used a graphic equalizer on a hi-fi system; however, the inverse of this relationship is not so widely appreciated.

An impulse in the time domain is a sinusoid in the frequency domain, or a shift operation in the time domain is complex modulation in the frequency domain. For example, a simple system that only delays the input by $m$ samples, has a frequency response of $H_m(\omega) = e^{-i\omega m}$. This relationship is understood by anybody who has designed an $M$-tap FIR filter by least-squares fitting the achieved frequency response to the desired frequency response. This process is also an (inverse) equalization problem as the filter's frequency response is simply a linear combination of sinusoids, with the amplitude of each sinusoid scaled by the coefficient at each filter tap, i.e. $H(\omega) = \sum_{k=-K}^{K} b[K + k]e^{-i\omega k}$.

This principle allows functional FIR filters to be designed very easily, without having to deal with the perplexities of feedback, for which the Laplace and $\mathcal{Z}$ transforms are ideally suited. Indeed, even the Fourier transform may be bypassed if time-domain design methods (e.g. based on weighted least-squares regression) are utilized [19],[20]. The details of FIR filter design will be covered in a future tutorial on detection, where 2-D filtering for spatial processing in imaging sensors will be treated. This is a more intuitive context for filter design because spatial filters may be analyzed and realized using delay and advance indexing in non-recursive FIR filters (i.e. spatial shifts), or forward and backward processing in recursive IIR filters [22],[23], without invoking notions of causality and non-causality, which only confound an already difficult subject.

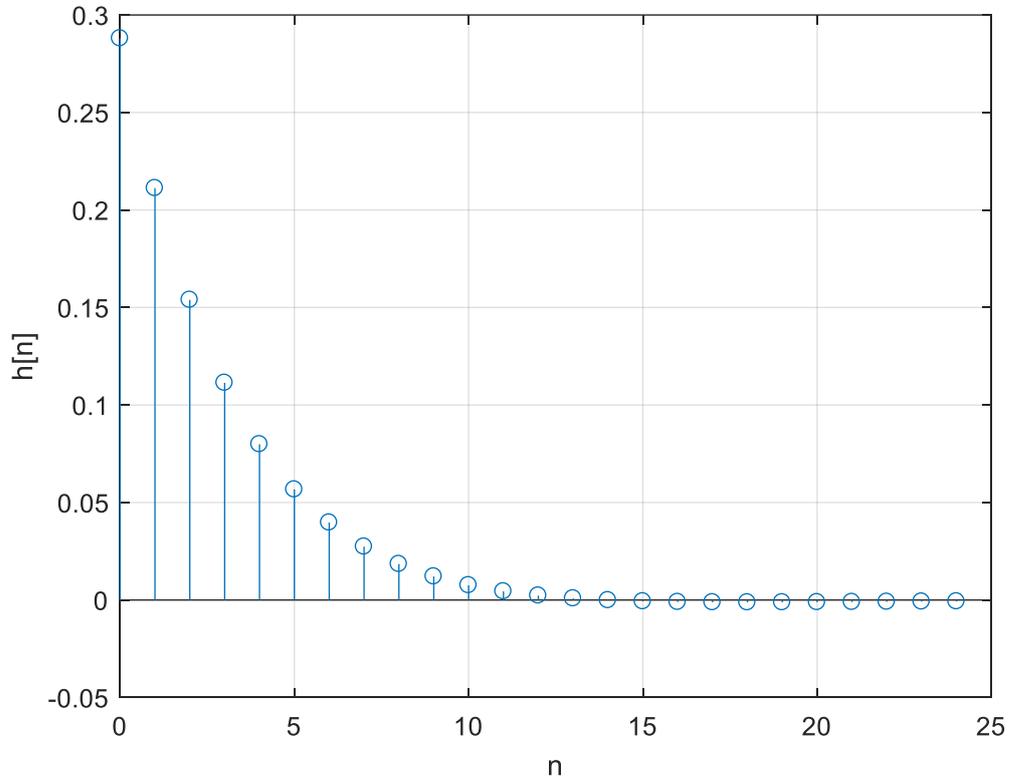

*Figure 13 - Impulse response for the $k = 0$ element (i.e. position) output by the state observer derived in Section 11, with all poles at $z = 0.8$ and an ideal group-delay of 2 samples.*

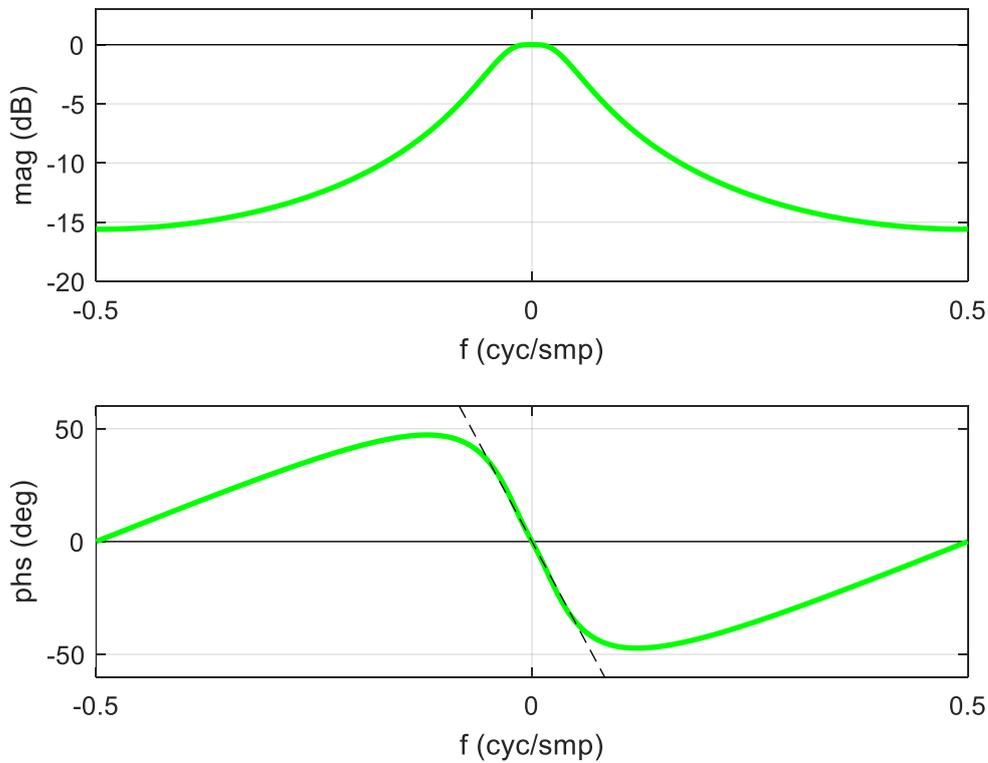

*Figure 14 – Frequency response as a function of normalized frequency ($f = \omega/2\pi$, cycles per sample) for the $k = 0$ element (i.e. position) output by the state observer derived in Section 11, with all poles at $z = 0.8$ and an ideal group-delay of 2 samples. Magnitude (dB) in upper subplot, phase (degrees) in lower subplot; both are evaluated from the complex frequency-response $H(\omega)$ shown in Figure 15. The ideal phase shift for the desired delay also shown (dashed black line).*

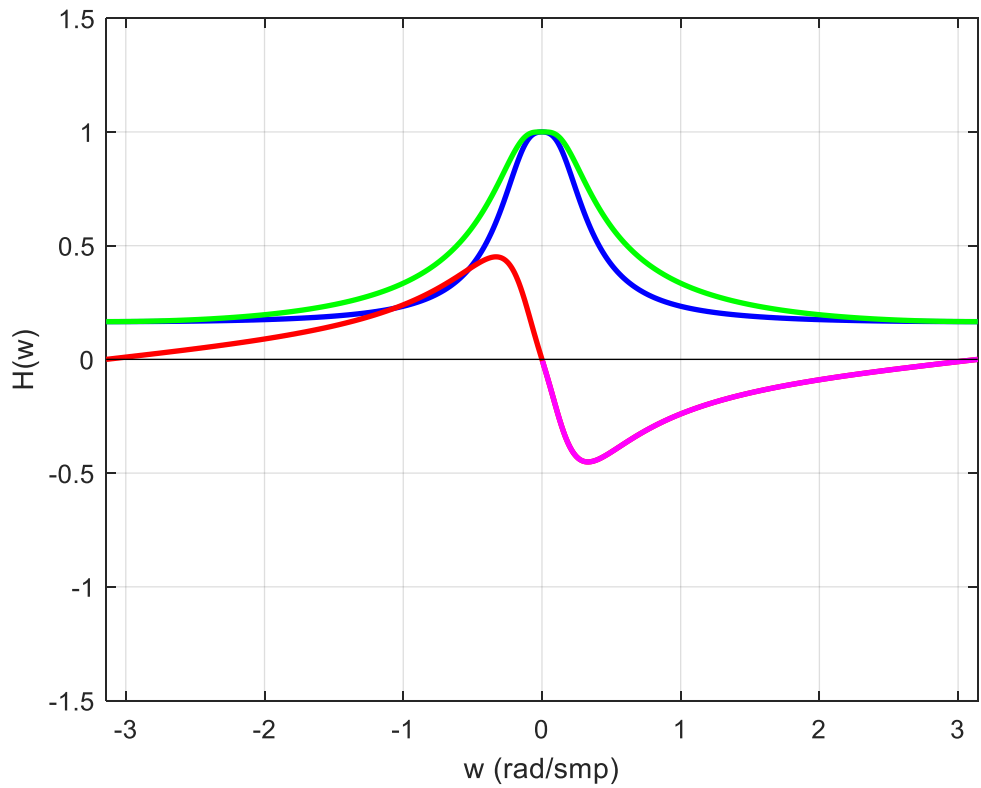

*Figure 15 – Frequency response $H(\omega)$, as a function of angular frequency (ω, radians per sample). Positive real part in blue, negative real part in cyan, positive imaginary part in red, negative imaginary part in magenta, magnitude in green. This complex response is used to evaluate the magnitude and phase in Figure 14.*

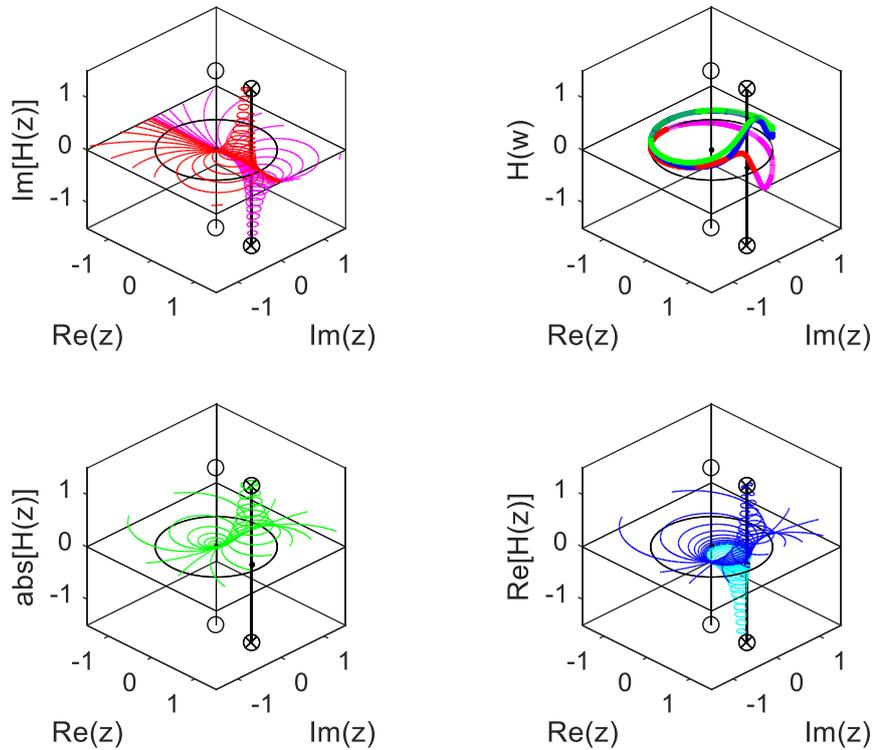

*Figure 16 – Discrete-time transfer-function $\mathcal{H}(z)$ evaluated in the complex z-plane. Positive real part in blue, negative real part in cyan, positive imaginary part in red, negative imaginary part in magenta, magnitude in green. The top-right subplot shows how $H(\omega)$ in Figure 15 is determined by evaluating $\mathcal{H}(z)$ around the unit circle. Locations of poles and zeros are shown using the '×' and '○' tokens, respectively. The filter poles boost the gain at low frequencies and cut the gain at high frequencies, because they are closest to the 'dc point' of the unit circle where $z = 1$ and furthest from the 'Nyquist point' of the unit circle, where $z = -1$. The filter zeros ensure that the frequency response has the required flatness (i.e. response derivatives w.r.t frequency) at dc, for the estimation of the selected kinematic state (i.e. an input derivative w.r.t time).*

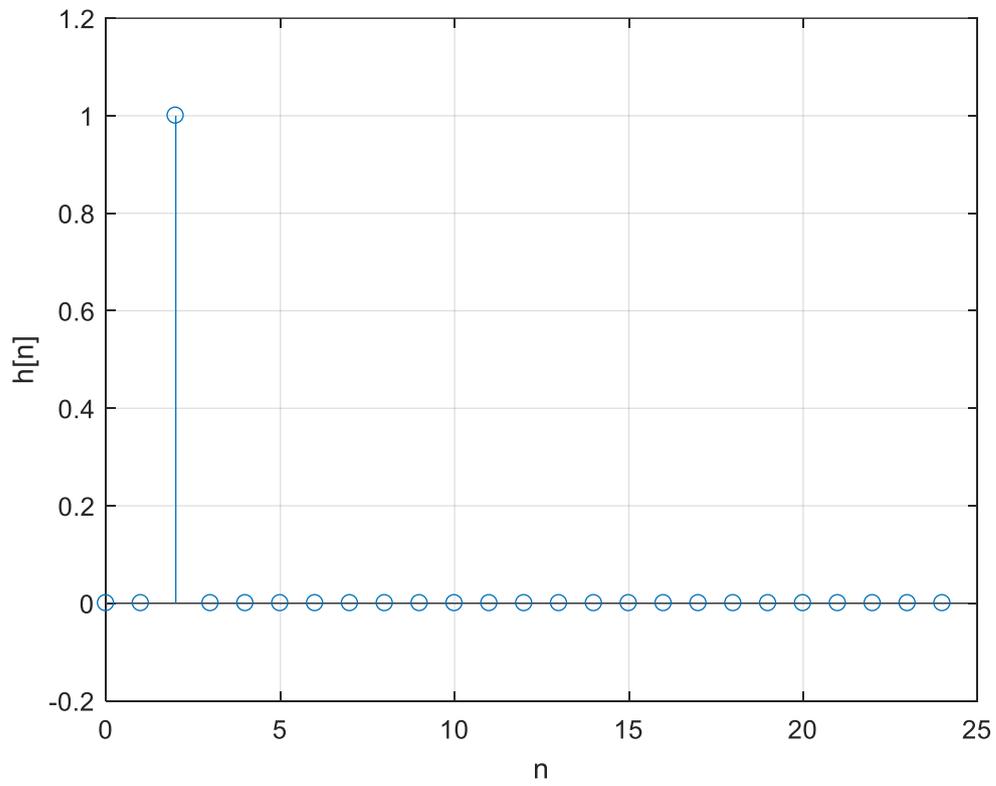

*Figure 17 - Impulse response for the $k = 0$ element (i.e. position) output by the state observer derived in Section 11, with all poles at $z = 0.0$ and group-delay of 2 samples.*

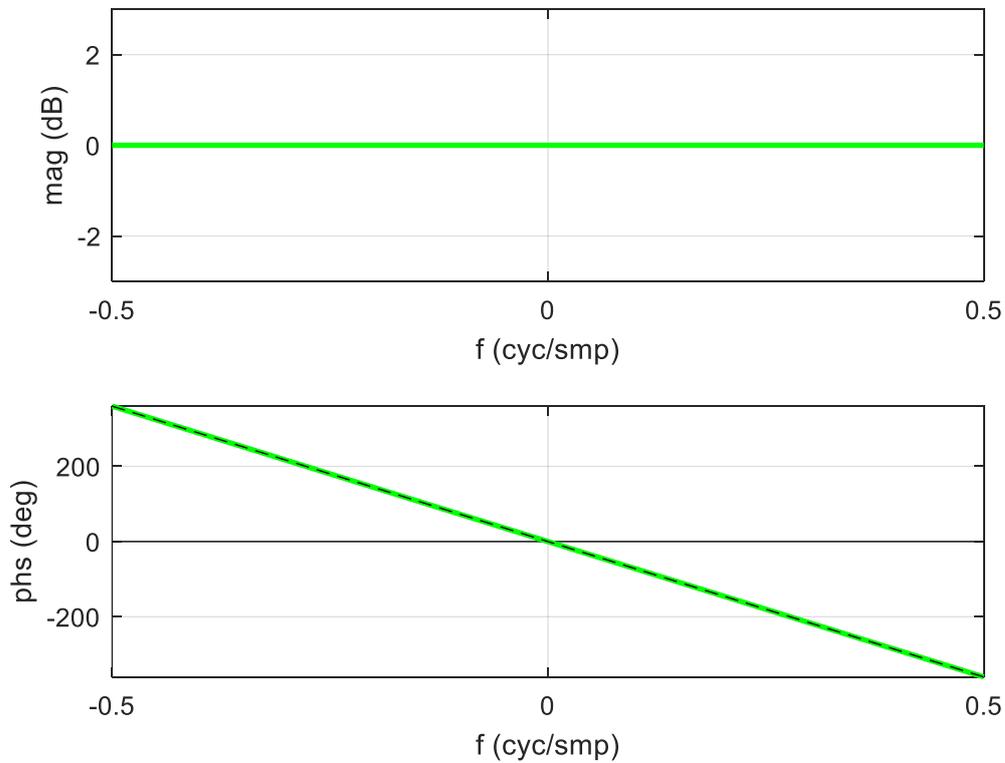

*Figure 18 – Frequency response as a function of normalized frequency ($f = \omega/2\pi$, cycles per sample) for the $k = 0$ element (i.e. position) output by the state observer derived in Section 11 with all poles at $z = 0.0$ and an ideal group-delay of 2 samples. Magnitude (dB) in upper subplot, phase (degrees) in lower subplot; both are evaluated from the complex frequency-response $H(\omega)$ shown in Figure 19. The ideal phase shift for desired delay is also shown (dashed black line).*

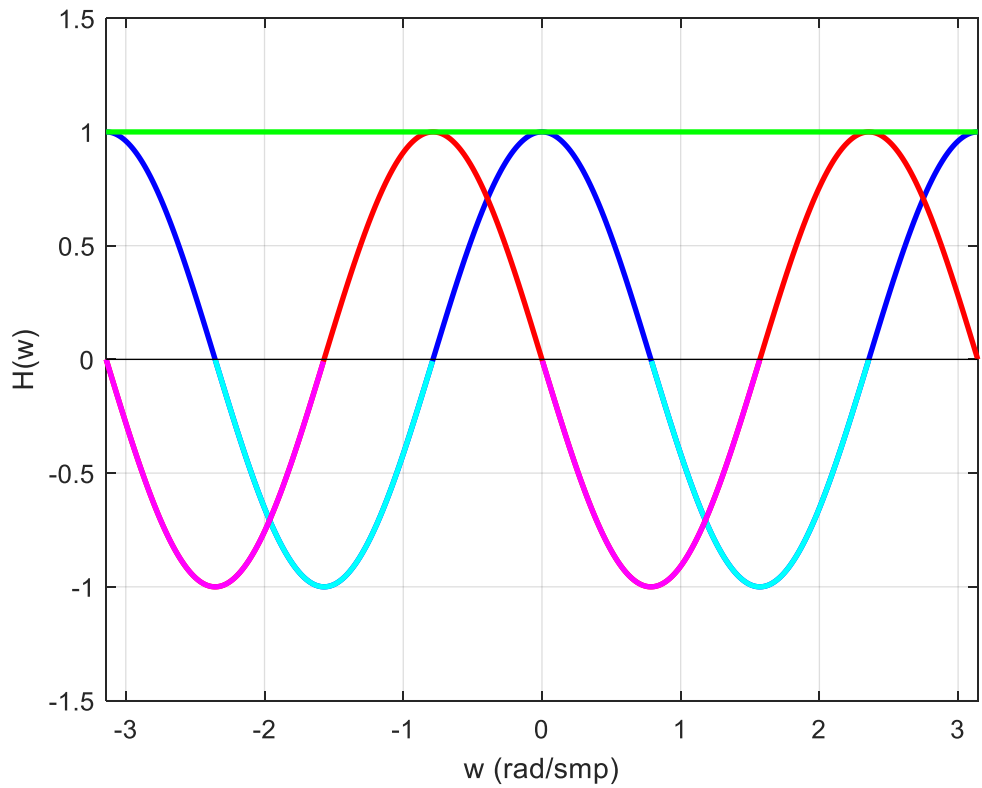

*Figure 19 – Frequency response $H(\omega)$, as a function of angular frequency (ω, radians per sample). Positive real part in blue, negative real part in cyan, positive imaginary part in red, negative imaginary part in magenta, magnitude in green. This complex response is used to evaluate the magnitude and phase in Figure 18.*

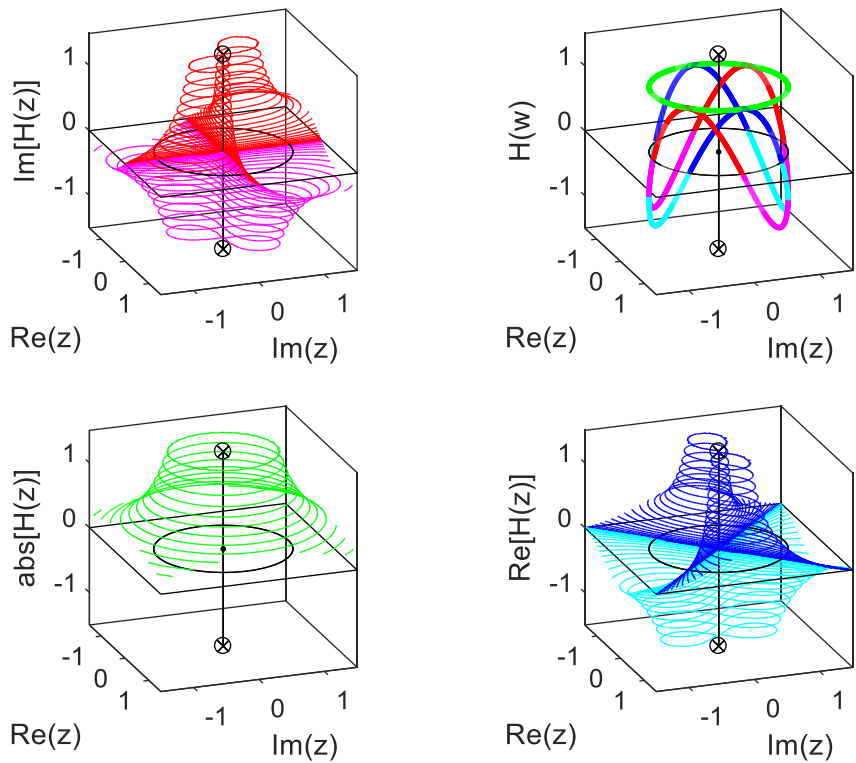

*Figure 20 – Discrete-time transfer function $\mathcal{H}(z)$ evaluated in the complex z-plane. Positive real part in blue, negative real part in cyan, positive imaginary part in red, negative imaginary part in magenta, magnitude in green. The top-right subplot shows how $H(\omega)$ in Figure 19 is determined by evaluating $\mathcal{H}(z)$ around the unit circle. Locations of poles and zeros are shown using the '×' and '○' tokens, respectively. All frequencies are affected equally (w.r.t. their phase and magnitude) because they are equidistant from the poles and zeros that are all at the centre of the unit circle where $z = 0$.*